\documentclass[trackchanges,twocolumn]{aastex7}

\usepackage{chngcntr}
\counterwithout{table}{section}

\begin{document}

\title{\textit{JWST} NIRSpec G395H Transmission Spectra of Two Similar Sub-Neptunes in the TOI-125 System} 

\author[0009-0008-5865-5831]{Yiwei Chai}
\affiliation{William H. Miller III Department of Physics and Astronomy, Johns Hopkins University, 3400 N. Charles Street, Baltimore, MD 21218, USA}
\email[show]{mchai3@jhu.edu}  

\author[0000-0001-9513-1449]{N\'{e}stor Espinoza} 
\affiliation{Space Telescope Science Institute, 3700 San Martin Drive, Baltimore, MD 21218, USA}
\affiliation{William H. Miller III Department of Physics and Astronomy, Johns Hopkins University, 3400 N. Charles Street, Baltimore, MD 21218, USA}
\email{nespinoza@stsci.edu}

\author[0000-0003-0652-2902]{Chloe Fisher} 
\affiliation{Astrophysics, Department of Physics, University of Oxford, Parks Road, Oxford, OX1 3RH, UK}
\email{chloe.fisher@physics.ox.ac.uk}

\author[0000-0002-2160-8782]{Erik Meier Valdes} 
\affiliation{Astrophysics, Department of Physics, University of Oxford, Parks Road, Oxford, OX1 3RH, UK}
\email{erik.meiervaldes@physics.ox.ac.uk}

\author[0000-0003-0030-332X]{Matthew Hooton} 
\affiliation{Cavendish Laboratory, JJ Thomson Avenue, Cambridge CB3 0HE, UK}
\email{mh2143@cam.ac.uk}

\author[0000-0002-4235-6369]{Brett Morris} 
\affiliation{Space Telescope Science Institute, 3700 San Martin Drive, Baltimore, MD 21218, USA}
\email{bmorris@stsci.edu}

\author[0000-0002-0832-710X]{Natalie H. Allen}
\affiliation{William H. Miller III Department of Physics and Astronomy, Johns Hopkins University, 3400 N. Charles Street, Baltimore, MD 21218, USA}
\email{nallen19@jhu.edu}

\author[0000-0002-6523-9536]{Adam J. Burgasser}
\affiliation{Department of Astronomy \& Astrophysics, University of California San Diego, La Jolla, CA 92093, USA}
\email{aburgasser@ucsd.edu}

\author[0000-0002-9355-5165]{Brice-Olivier Demory}
\affiliation{Center for Space and Habitability, University of Bern, Bern, Switzerland}
\affiliation{ARTORG Center for Biomedical Engineering Research, University of Bern, Bern, Switzerland}
\affiliation{Space Research and Planetary Sciences, Physics Institute, University of Bern, Bern, Switzerland}
\email{brice-olivier.demory@unibe.ch}

\author[0000-0003-0854-3002]{Am\'{e}lie Gressier}
\affiliation{Department of Physics and Trottier Institute for Research on Exoplanets, Universit\'{e} de Montr\'{e}al, Montr\'{e}al, QC, Canada}
\email{amelie.gressier@umontreal.ca}

\author[0000-0003-1907-5910]{Kevin Heng}
\affiliation{Faculty of Physics, Ludwig Maximilian University, Scheinerstrasse 1, D-81679, Munich, Bavaria, Germany}
\affiliation{Munich Center for Geoastronomy, Ludwig Maximilian University, Theresienstrasse 41, D-80333, Munich, Bavaria, Germany}
\affiliation{University College London, Department of Physics \& Astronomy, Gower St, London, WC1E 6BT, UK}
\email{kevin.heng@physik.lmu.de}

\author[0000-0003-3204-8183]{Mercedes L\'{o}pez-Morales}
\affiliation{Space Telescope Science Institute, 3700 San Martin Drive, Baltimore, MD 21218, USA}
\email{mlopez-morales@stsci.edu}

\author[0000-0002-7384-8577]{Meng Tian}
\affiliation{Faculty of Physics, Ludwig Maximilian University, Scheinerstrasse 1, D-81679, Munich, Bavaria, Germany}
\email{meng.tian@physik.lmu.de}

\begin{abstract}

The \textit{James Webb Space Telescope (JWST)} has enabled the first in-depth characterisations of sub-Neptune atmospheres, revealing a diversity of chemical compositions and cloud properties that challenge current models. Multi-planet systems provide an opportunity to compare planetary atmospheres while controlling for formation conditions within the same protoplanetary disk. We present \textit{JWST} NIRSpec G395H 3--5 $\mu m$ observations of two similarly-sized sub-Neptunes orbiting a sun-like star within the TOI-125 system: TOI-125 b ($P=4.65$ d, $R_p=2.7~R_\oplus$, $M_p=10~M_\oplus$, $T_\mathrm{eq}=1100$ K) and TOI-125 c ($P=9.15$ d, $R_p=2.85~R_\oplus$, $M_p=6.8~M_\oplus$, $T_\mathrm{eq}=825$ K), comprising of one full transit of TOI-125 b but only 70\% of TOI-125 c's transit due to larger than predicted TTVs. We obtain precise $R\sim100$ transmission spectra for both planets, with median errorbars of 50 ppm for TOI-125 b and 85 ppm for TOI-125 c. Atmospheric retrievals are compatible with a $\sim$150x solar metallicity atmosphere or clouds above $\sim$3 mbar for TOI-125 b, and a $\sim$100x solar metallicity atmosphere or clouds above $\sim$10 mbar for TOI-125 c. Our results may support recent evidence for metal enrichment in sub-Neptunes, while also further complicating the proposed trend in atmospheric feature size with planet temperature. However, shorter wavelength follow-up observations of both planets are needed to establish whether their spectra are truly muted or feature-rich, with implications for the nature of sub-Neptunes around sun-like stars.
\end{abstract}


\section{Introduction}

The \textit{Kepler} mission established that sub-Neptunes ($2R_\oplus < R < 4R_\oplus$) are the most common type of exoplanet in our local neighbourhood \citep[for planets with $P<100$ d; see][and references therein]{Batalha2013}. Unlike terrestrial or giant planets, sub-Neptunes have no clear analogue in the Solar System; this poses a challenge to our current planet formation and evolution models and makes their composition and origin key open questions in exoplanet science. Atmospheric studies via transmission spectroscopy have emerged as a powerful method for characterising sub-Neptunes in order to probe their composition, structure and even formation histories. However, such observations are challenging because of the planets' small relative size compared to their host stars, resulting in small measured differences in transit depth as a function of wavelength.

Currently, \textit{Hubble Space Telescope (HST)} and the \textit{James Webb Space Telescope (JWST)} have driven most detections of individual absorbers in sub-Neptune atmospheres, due to their ability to provide the photon-collecting area and spectrophotometric stability needed to resolve the small ($\sim$10s--100s of ppm), broadband molecular features expected in sub-Neptune transmission spectra. In more recent years, high-resolution ground-based facilities such as CRIRES, CARMENES and WINERED have also sought to place constraints on sub-Neptune atmospheres, particularly regarding signatures of atmospheric escape \citep[see, e.g.,][]{OrellMiquel2023_TOI1430b_CARMENES_He_escape, Zhang2025_TOI836c_Keck_He_escape, Vissapragada2026_subNep_GJ3090b_WINRED_He_escape}, although many have yielded only weak constraints or upper limits on detections of other molecular signatures \citep[see, e.g.,][]{Dash2024_GJ3470b_CARMENES_H2O_upperlimit, Parker2025_CRIRES_subNep_GJ3090b_nondetect, Nortmann2026_GJ1214b_CO2_Kband_CRIRES}.

From the initial \textit{HST} studies, sub-Neptunes were revealed to possess a range of feature-rich to muted or featureless spectra, hinting at a diversity of atmospheric conditions \citep[e.g.][]{Kreidberg2014_GJ1214b, Fraine2014_HatP11b, Wakeford2017_HAT-P-26b, Benneke2019_K218b, Mikal-Evans2023_HST_TOI270d}. \textit{JWST}, which has enabled significant progress in both sub-Neptune and other exoplanet science \citep[see][for a review]{Review_Espinoza2025_JWSTexohighlights}, has further unveiled a varied molecular inventory amongst planets with feature-rich spectra. Temperate sub-Neptunes TOI-270 d \citep{Benneke2024arXiv_miscible_subNeptunes, holmberg2024_toi270d_nirspec, felix2025A_toi270d_niriss} and K2-18 b \citep{Madhusudhan2023_K218b_NIRISS_NIRSpec} are dominated by CH$_4$ absorption and show observable features due to CO$_2$ and perhaps CS$_2$ --- although, interestingly, LP 791-18 c, which lies between K2-18 b and TOI-270 d in radius, mass and equilibrium temperature ($T_\mathrm{eq}$), shows no sign of CO$_2$ and has a spectrum dominated by haze scattering \citep{Roy2026}. For GJ 9827 d \citep{piauletghorayeb2024_GJ9827d_nirisss} and TOI-421 b \citep{Davenport2025}, H$_2$O is the dominant absorber, while SO$_2$ absorption, indicative of photochemistry, has been detected in GJ 3470 b \citep{Beatty2024_GJ3470b_SO2}. 

There is also a substantial fraction of observed sub-Neptunes which exhibit flat or muted transmission spectra, e.g. TOI-836 c \citep{Wallack2024_COMPASS_TOI836c}, GJ 1214 b \citep{Kreidberg2014_GJ1214b, Schlawin2024a_GJ1214b_NIRSpec}, TOI-776 c \citep{Teske2025_COMPASS_TOI776c}, HD 15337 c \citep{Wallack2026_COMPASS_HD15337c}. This has been predicted to be either a result of high mean molecular weight (MMW) atmospheres \citep{Miller-Ricci2009} or obscuration by clouds and hazes \citep{Benneke&Seager2013, Morley2013}. However, for most of these planets, there still remains a degeneracy as to which of these effects dominates. This may be due to the choice of wavelength coverage; many of these planets were observed from 3--5 $\mu m$ with \textit{JWST} NIRSpec/G395H, an instrument mode popularly used for transmission spectroscopy due to its higher resolution and coverage of a wavelength region containing absorption features from key molecules such as H$_2$O, CH$_4$, CO, CO$_2$ and SO$_2$. 

These data were mostly taken during earlier \textit{JWST} observing cycles, at a time when there was no actual empirical on-sky data available that could be transferred to the \texttt{PandExo} ETC \citep{batalha2017_pandexo}. This made it more difficult to accurately estimate the noise properties of simulated data and thus the observing requirements to confidently detect particular molecular features in the spectra. ndeed, learned empirical error bars from early on-sky JWST observations revealed that the original \texttt{PandExo} error bars from \cite{batalha2017_pandexo} were slightly underestimated.

Since then, it has been suggested that shorter wavelength data (e.g. with NIRISS/SOSS) may be useful for breaking the cloud--metallicity degeneracy. \cite{Davenport2025} highlight that, had they only observed TOI-421 b from 3--5 $\mu m$, they would have incorrectly concluded that it possessed a featureless spectrum; conversely, their 1--3 $\mu m$ NIRISS/SOSS measurements alone are effective for constraining the planet's atmospheric water abundance and MMW. This is likely because these wavelengths cover a region of prominent water absorption features, compared to the relatively smaller or more muddled signatures of molecules in the 3--5 $\mu m$ region. Indeed, the usefulness of shorter wavelength observations has even been shown in the high-metallicity regime where features are overall expected to be more muted; \cite{piauletghorayeb2024_GJ9827d_nirisss} find that NIRISS data combined with HST/WFC3 enabled the measurement of water in the warm, high-MMW sub-Neptune GJ 9827 d.  

Naturally, the observed diversity of sub-Neptune spectra has spurred comparative studies in the attempt to interpret its underlying mechanisms. Notably, \cite{Brande2024} have suggested a potential parabolic trend in 1.4 $\mu m$ atmospheric feature size with $T_\mathrm{eq}$, in which planets in the $\sim$500--850 K range often exhibit muted or featureless spectra, while those at cooler and hotter temperatures typically show spectral features. This has been attributed to the impact of atmospheric chemistry processes and/or cloud formation: in the $\sim$500--850 K regime, the dominant carbon-bearing species is thought to be methane, which tends to drive production of photochemical hydrocarbon hazes that in turn obscure spectral features \citep{Morley2015, Gao2020_silicatesabove950_hydrocarbonsbelow}. At temperatures below $\sim500$ K, rainout may lead to clearer atmospheres \citep{Brande2024}, whereas at temperatures above $\sim$850 K, carbon monoxide replaces methane as the dominant carbon-bearing species, resulting in reduced haze production.

These findings are, however, based on a small sample of sub-Neptunes with $T_\mathrm{eq}<$1000 K, the vast majority of which orbit M- and K-dwarfs. Additionally, many of these planets, e.g. K2-18 b and TOI-270 d, may possess relatively high MMW ($\mu>5$ amu) atmospheres, which complicates the low-MMW bulk H$_2$/He model ($\mu\sim2.3$ amu) initially theorised from the \textit{Kepler} population of sub-Neptunes. This may indicate a need for new classifications such as water-worlds \citep[i.e. bulk compositions of $\sim$10$\%$ H$_2$O;][]{Madhusudhan2021_Hycean, Hu2021_oceanworlds} or miscible-envelope planets \citep[i.e. mixed H$_2$/H$_2$O outer envelopes;][]{Benneke2024arXiv_miscible_subNeptunes} --- although see \cite{Heng2025_radiusvalley_MMW_gradient}, who predict that sub-Neptunes with higher MMWs may be the result of geochemical outgassing, which more strongly affects the atmospheric composition of smaller planets less able to hold onto their primordial H/He envelopes.

Given that hazes are thought to be dependent on the incident spectrum of the host star \citep{Arney2017}, it is important to control for host star type when looking for aerosol trends in exoplanet atmospheres. Recognising the need for more data across a broader range of host stars and planetary parameters, recent studies have begun to probe sub-Neptune atmospheres at hotter equilibrium temperature regimes (i.e. $>800$ K) around sun-like stars. \textit{JWST} NIRISS SOSS $+$ NIRSpec G395M 1--5 $\mu m$ observations of TOI-421 b \citep[$P=5.19$ d, $R_p=2.64~R_\oplus$, $M_p=6.7~M_\oplus$, $T_\mathrm{eq}=920$ K;][]{Davenport2025} and \textit{HST} WFC3 $+$ STIS 1.1--1.7 $\mu m$ observations of HD 86266 c \citep[$P=3.98$ d, $R_p=2.16~R_\oplus$, $M_p=7.25~M_\oplus$, $T_\mathrm{eq}=1310$ K;][]{Kahle2025}, which both orbit G-type hosts, have sought to test whether sub-Neptunes orbiting sun-like stars follow the temperature-dependent trend in spectral feature size proposed by \cite{Brande2024} for sub-Neptunes around M- and K-dwarfs. For TOI-421 b, which falls above the temperature regime at which photochemical hazes are expected to obscure spectral features, \cite{Davenport2025} report a feature-rich transmission spectrum consistent with a clear, low-MMW atmosphere. This is in keeping with the parabolic trend measured from M-dwarf sub-Neptunes. However, \cite{Kahle2025} find that HD 86266 c (which has a significantly higher $T_\mathrm{eq}=1300$ K), presents a featureless spectrum, with retrieval results ruling out a clear, solar-composition atmosphere. One proposed explanation for HD 86226 c's lack of spectral features at $\sim1.4~\mu m$ is that the onset of silicate cloud formation, which is thought to occur at $T_\mathrm{eq} > 950$ K, may be obscuring atmospheric features at these higher temperature regimes.

Also of note is the recent \textit{JWST} NIRISS SOSS $+$ NIRSpec G395H 1--5 $\mu m$ observation of TOI-1130 b \citep[$P=4.07$ d, $R=3.66 R_\oplus$, $M=19.8 M_\oplus$, and $T_\mathrm{eq}=825$ K;][]{Barat2026_TOI-1130b}. While perhaps closer to Neptune in size and mass than to the typical sub-Neptune, and orbiting a K7-type dwarf, its high $T_\mathrm{eq}$ places it in a similar temperature regime to TOI-421 b. Interestingly, it too shows a feature-rich spectra dominated by H$_2$O features, but has $\mu=5.5$ amu, similar to that of TOI-270 d and potentially consistent with a miscible envelope scenario.

Ultimately, the continued observations of spectral diversity in sub-Neptunes around both sun-like stars and M-dwarfs raises questions about whether the atmospheres of objects in different regions of the sub-Neptune parameter space have fundamentally different natures. To this end, we report spectroscopic \textit{JWST} observations of TOI-125 b and TOI-125 c, two similar sub-Neptunes within a single system orbiting the K0V star TOI-125 \citep[$T_\mathrm{eff}=5320\pm39$ K, $M=0.859^{+0.044}_{-0.038}~M_\odot$, $R=0.84\pm0.011~R_\odot$;][]{Nielson2020_TOI125_masses}. The inner planet b hasperiod $P=4.65\pm0.00004$ d, radius $R=2.7\pm0.03~R_\oplus$, mass $M_p=10^{+0.9}_{-0.8}~M_\oplus$ and equilibrium temperature $T_\mathrm{eq}=1100^{+35}_{-38}$ K. The outer planet c has period $P=9.15\pm0.00001$ d, radius $R=2.85\pm0.04~R_\oplus$, mass $M_p=6.8\pm0.8~M_\oplus$ and equilibrium temperature $T_\mathrm{eq}=825^{+45}_{-48}$ K (this work; see Appendix \ref{sec:appA}). Additionally, the system has a third confirmed planet, TOI-125 d, with $P=19.98\pm0.00003$ d, $R=2.45^{+14}_{-12}~R_\oplus$, $M=13.9\pm1.0~M_\oplus$ and $T_\mathrm{eq}=580^{+35}_{-43}$ K (this work), although it was not observed as it was predicted to have a smaller atmospheric scale height and less pronounced spectral features due to its cooler temperature.

We present the first transmission spectra for planets b and c, obtained from 3--5 $\mu m $ with \textit{JWST} NIRSpec G395H at $R\sim2700$. In Section \ref{sec:methods}, we describe the \textit{JWST} NIRSpec observations and our methods for the data reduction, light curve fits, and atmospheric retrievals. In Section \ref{sec:results}, we present our analysis of the resulting transmission spectrum and retrieved atmospheric properties for each planet. Section \ref{sec:discuss} discusses the implications of our results for both planets in the TOI-125 system and for the wider sub-Neptune population, as well as possibilities for follow-up observations. We summarise our conclusions in Section \ref{sec:concl}.

\section{Methods}\label{sec:methods}

\subsection{Observations}

We observed TOI-125 b and c using the \textit{JWST} NIRSpec Bright Object Time Series (BOTS) mode as part of Cycle 2 program \#4126 \citep{Fisher2023_TOI125_JWST_PID4126F}. Observations targeted a single primary transit of each planet, with target acquisition for each visit performed using the host star TOI-125 ($m_J=9.466\pm0.021$). TOI-125 b was observed on September 30, 2024 and TOI-125 c was observed on November 27, 2024. Observations for both planets consisted of a single exposure with the G395H/F290LP disperser/filter combination and the SUB2048 subarray in the NRSRAPID readout pattern. This provided wavelength coverage from 2.8--5.14 $\mu m$ across detectors NRS1 and NRS2, excepting a 3.72--3.82 $\mu m$ gap between the two. Each exposure lasted 7.66 hours and used 22 groups/integration for a total of 1328 integrations per exposure. We note that only the first 70\% of the transit of TOI-125 c was captured due to larger-than-anticipated transit timing variations (TTVs; see Appendix \ref{sec:appA} for details).

\subsection{Data Reduction}

To test the robustness of our transmission spectra, we perform two independent reductions: one with \texttt{transitspectroscopy} \citep{transitspectroscopy} and one with \texttt{Eureka!} \citep{Bell2022_Eureka}. We make use of the spectra produced by both reductions in the following parts of this paper; as such, we provide a description of the reduction steps for each method below. We define columns as the pixels along the wavelength axis of the spectra, and rows as the pixels along the cross-dispersion axis.

\subsubsection{\texttt{transitspectroscopy} Reduction}

The NRS1 and NRS2 data were reduced separately using the \texttt{transitspectroscopy} pipeline version 0.4.1 \citep{transitspectroscopy}, which in turn makes use of the \textit{JWST} pipeline version 1.17.1 and CRDS context \texttt{jwst$\_$1322.pmap}. The pipeline begins with the uncalibrated data products for all observations in the form of \texttt{.uncal} files. We perform data quality initialisation, saturation checks, superbias subtraction, reference pixel correction, linearity correction and dark subtraction using the default settings from Stage 1 of the \textit{JWST} pipeline; however, we do not apply 1/f noise correction at this stage. \texttt{transitspectroscopy} also performs a customised TSO jump detection algorithm as follows: for each integration, we calculate the difference in fluence from groups $i$ to $i+1$ to obtain a `group difference'. A median filter with a window of 10 integrations is then subtracted from this group difference time series, leaving only the noise on the time series. Any values deviating more than 10-$\sigma$ from this noise are flagged as jumps for that group and assigned a \texttt{groupdq} value of 4. This process is repeated from group $i=1$ to group $N_\mathrm{groups}=i-1$. After the custom jump-detection step, we obtain the rates per integration using the default ramp-fitting algorithm from the \textit{JWST} pipeline. This produces uncalibrated slope images in the form of \texttt{.rateints} files.

To extract the spectrum from these rates per integration, \texttt{transitspectroscopy} performs cross-correlation with a Gaussian of width 1-pixel at each column to obtain the centre of the trace. The resulting curved trace is then smoothed with a B-spline using 8 equally spaced knots. 1/f noise correction is applied at the ramp-level similar to the method used in \cite{Radica2023} by removing the scaled median frame from each ramp, estimating the 1/f noise from the resulting frame, and then removing it from the original frame. Spectral extraction is performed via simple extraction along the centre of the smoothed trace for each integration. We test extraction apertures from 2--12 pixels in size and find that the 2-pixel aperture produces the smallest RMS on the resulting white-light light-curve for both NRS1 and NRS2. As such, we use the 2-pixel aperture when extracting the spectroscopic light curves. Finally, outliers on the spectroscopic light curves are removed by clipping points over a 5-$\sigma$ threshold from the median absolute deviation of a 15-point wide median filter.

\subsubsection{\texttt{Eureka!} Reduction}

The data were reduced separately for NRS1 and NRS2 using version 1.2 of the \texttt{Eureka!} pipeline \citep{Bell2022_Eureka}, which acts as a wrapper around the \textit{JWST} pipeline; in this case, the \textit{JWST} pipeline version was 1.18.0 with CRDS context \texttt{jwst$\_$1364.pmap}.

\begin{figure*}[!ht]
    \centering
    \includegraphics[width=\textwidth]{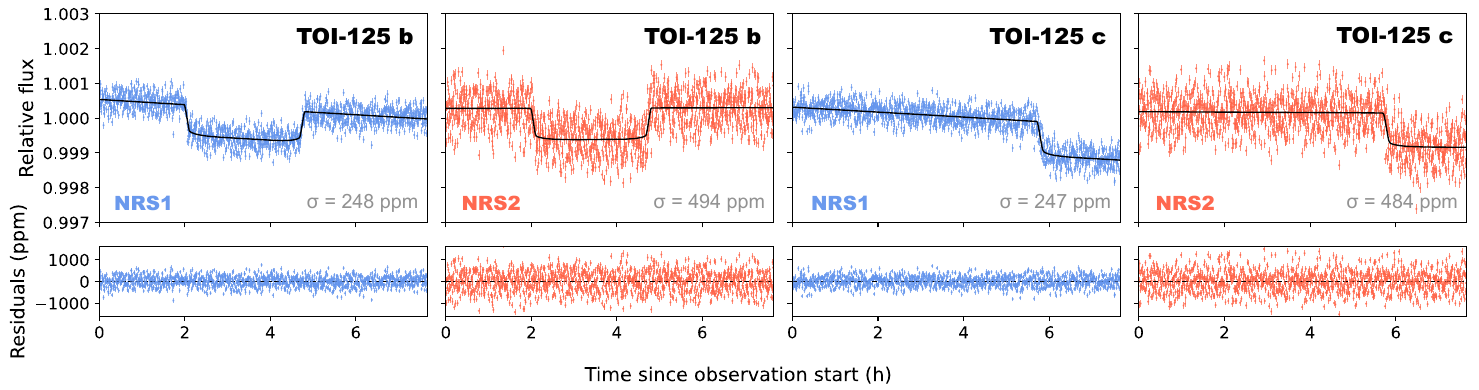}
\end{figure*}

\begin{figure*}[!ht]
    \centering
    \includegraphics[width=\textwidth]{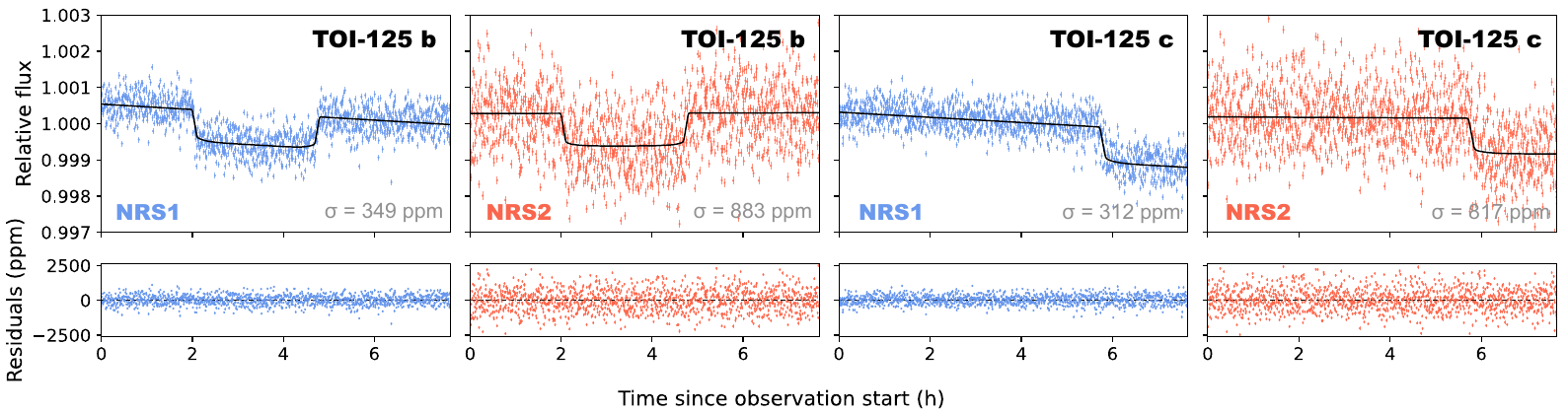}
    \caption{White-light light-curves from the \texttt{transitspectroscopy} (top) and \texttt{Eureka!} (bottom) reductions. Upper panels for each reduction show the normalised white-light light-curve fits to the NRS1 (blue) and NRS2 (orange) data for TOI-125 b and TOI-125 c. 
    Lower panels show residuals between the observed data and the best-fit planet transit model. 
    }
    \label{fig:1}
\end{figure*}

The reduction begins with the uncalibrated data products for all observations in the form of \texttt{.uncal} files. Detector-level calibration follows the same set-up as for the \texttt{transitspectroscopy} reduction, with two exceptions: both 1/f noise correction and jump detection are performed at this stage using the \textit{JWST} pipeline defaults. Following detector-level calibration, the position of the spectral trace is determined from the output rates per integration by modelling a Gaussian within a half-width of 4 pixels. The frames are trimmed outside of columns 600 and 2043 and rows 1 to 30 for NRS1, and columns 1 and 2044 and rows 1 and 30 for NRS2. Trace curvature is corrected and the background is subtracted column by column at a half-width of 7 pixels. Spectral extraction is performed via optimal extraction using a 3-pixel aperture from the centre of the trace. Finally, to eliminate outliers from the resulting spectroscopic light curves, light curve points over 4-$\sigma$ from a rolling median with a width of 13 points are clipped.

\subsection{Light-Curve Fitting}

We first fit the white-light light-curves for both planets as part of a global fit on all available data, with the aim of calculating updated parameters for the TOI-125 system. The \textit{JWST} white-light light-curves for each planet are obtained by dividing the sum of all light-curves across instrument pixel-resolution wavelengths by the median of all such light-curves. Errors on the white-light light-curves were calculated by taking the quadrature sum of the spectroscopic light curve uncertainties at each timestamp.

Figure \ref{fig:1} shows the NRS1 and NRS2 white-light light-curve data for both planets, the best-fit model from the white-light light-curve fit and the associated residuals for the \texttt{transitspectroscopy} and \texttt{Eureka!} reductions. We find that the \texttt{transitspectroscopy} achieves a smaller root mean scatter of 248 ppm (NRS1) and 494 ppm (NRS2) for TOI-125 b and 247 ppm (NRS1) and 484 ppm (NRS2) for TOI-125 c. In comparison, the \texttt{Eureka!} reduction white-light light curve has RMS of 349 ppm (NRS1) and 883 ppm (NRS2) for TOI-125 b and 312 ppm (NRS1) and 817 ppm (NRS2) for TOI-125 c. This difference may be due to the choice in size of extraction aperture (2 vs 3 pixels respectively); when using a 3-pixel aperture with \texttt{transitspectroscopy}, we find a higher resulting white-light light curve RMS (see Section \ref{sec:methods}).

In our global fit, we elect to use the NRS1 and NRS2 white-light light-curves for both planets from the \texttt{transitspectroscopy} reduction due to the lower RMS, along with publicly available light-curves from TESS sectors 1, 2, 28, 68, 69 and HARPS radial velocities \citep{Nielson2020_TOI125_masses}. The global fit is performed with \texttt{juliet} version 2.2.8 \citep{Espinoza2019_juliet} using the \texttt{dynamic-dynesty} Nested Sampler \citep{Speagle2020_dynesty, Koposov2025_dynesty_v3.0.0} with 3000 live points. Further details for this global fit are provided in Appendix \ref{sec:appA}, along with the posterior distributions of the fitted parameters. 

\begin{figure*}[!t]
    \centering
    \includegraphics[width=\linewidth]{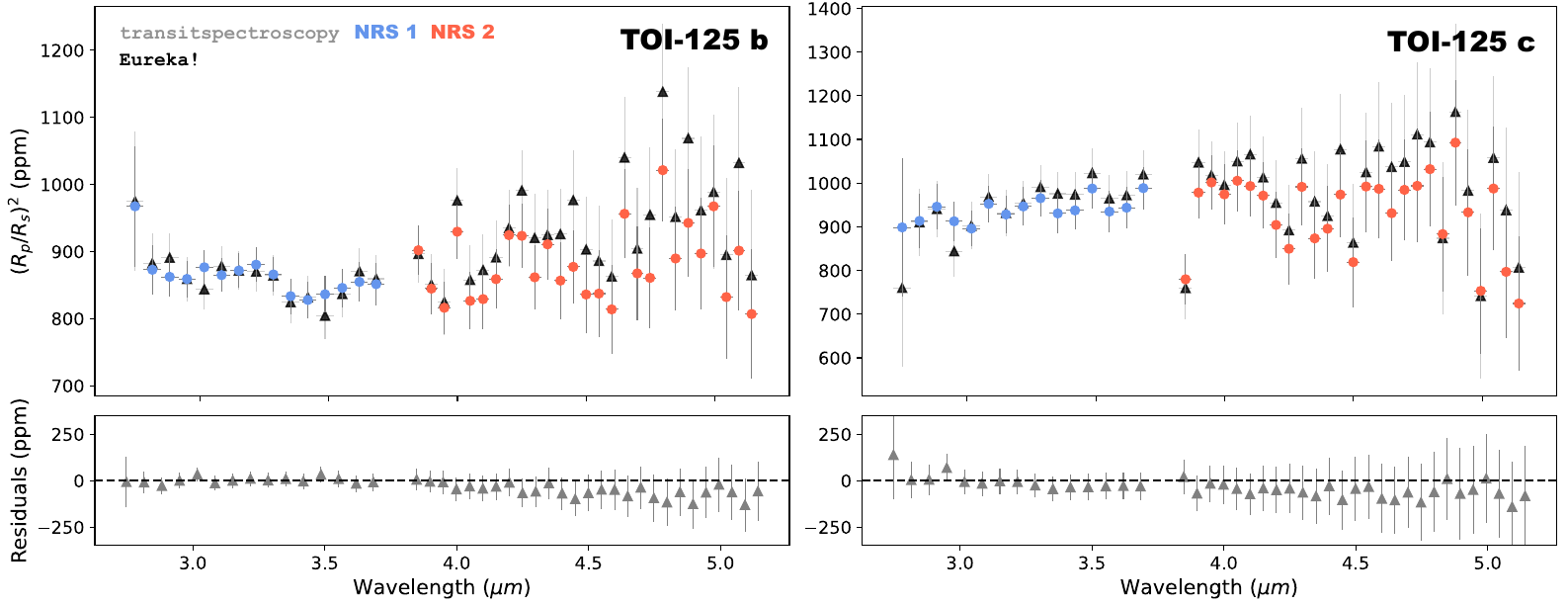}
    \caption{Comparison of the transmission spectra resulting from the independent \texttt{transitspectroscopy} (colour) and \texttt{Eureka!} (black) reductions. Top panels show the transmission spectra and lower panels show the residuals between the two reductions. The data and uncertainties included in this figure are available.}
    \label{fig:2}
\end{figure*}

To obtain the transmission spectrum of each planet, we apply the same method to both data reductions, as follows. The pixel-resolution light curves are binned into 44 wavelength bins of equal size across NRS1 $+$ NRS2, so that the wavelength-dependent spectroscopic light-curve fits are performed on these binned light curves. We fix values for the period ($P$), time of transit centre ($t_0$), impact parameter ($b$), normalised semi-major axis ($a/R_*$), eccentricity ($e$), and argument of periastron ($\omega$) according to the best-fit values from our global fit. We set a truncated normal prior on the quadratic limb-darkening coefficients, $u_1$ and $u_2$. The mean values for the distribution, $\mu_{u_1}$ and $\mu_{u_2}$, are obtained by calculating the expected $u_1,u_2$ values for each binned wavelength with \texttt{ExoTic-LD} \citep{Grant&Wakeford2024_exotic-ldc}. We use the MPS-ATLAS-2 stellar grid \citep{Kostogryz2022_MPS2} and input TOI-125 stellar parameters for the effective temperature ($T_\mathrm{eff}=5320$ K), surface gravity (log$(g)=4.516$) and metallicity ($-0.2$ dex) from \cite{Nielson2020_TOI125_masses}. Finally, we include a jitter term, $\sigma_\omega$, which is added in quadrature to the error-bars for each instrument. Table \ref{tab:1} reports the full list of priors used for the spectroscopic light curve fits. We obtain a transmission spectrum with resolution of $R\sim100$ for both planets by taking the resulting median posterior $R_p/R_*$ value from each spectroscopic light curve fit (see Figure \ref{fig:2}).

\begin{table}
    \begin{center}
        \caption{Priors on the Spectroscopic Light Curve Fits}
        \label{tab:1}
        \begin{tabular}{lll}
            \hline
            Parameter&  TOI-125 b&  TOI-125 c\\
            \hline
            $P$ (d)&  $4.65$&  $9.15$\\
            $T_0$ (BJD)&  $2458327.450$&  $2458325.289$\\
            $R_p/R_*$&  $U(-0.1,0.1)$&  $U(-0.1,0.1)$\\
            $b$&  $0.39$&  $0.56$\\
            $a/R_*$&  $13.2$&  $20.6$\\
            $e$&  $0.19$&  $0.04$\\
            $\omega$ (deg)&  $-32.5$&  $39.6$\\
            $u_1$&  t$N(\mu_{u_1}, 0.1, -3.0, 3.0)$&  t$N(u_1, 0.1, -3.0, 3.0)$\\
            $u_2$&  t$N(\mu_{u_2}, 0.1, -3.0, 3.0)$&  t$N(u_2, 0.1, -3.0, 3.0)$\\
            $M_\mathrm{flux}$&  $N(0, 0.001)$&  $N(0, 0.001)$\\
            $\sigma_\omega$ (ppm)&  log$U(1, 10000)$&  log$U(1, 10000)$\\
            $\theta_0$&  $U(-100,100)$&  $U(-100,100)$\\
            \hline
        \end{tabular}
    \end{center}
    \vspace{-2mm}
    \footnotesize{Fixed values are based on median posteriors of the global fit performed in this work (see Appendix \ref{sec:appA}). For the truncated normal prior on the quadratic limb darkening coefficients, mean values $\mu_{u_1}$ and $\mu_{u_2}$ are calculated at each binned wavelength using \texttt{ExoTiC-LD} \citep{Grant&Wakeford2024_exotic-ldc} with the MPS-ATLAS-2 stellar grid \citep{Kostogryz2022_MPS2}; inputs are TOI-125 stellar effective temperature $T_\mathrm{eff}=5320$ K, surface gravity log$(g)=4.516$ and metallicity $=-0.2$ dex from \cite{Nielson2020_TOI125_masses}.}
\end{table}

\subsection{Atmospheric Retrievals}

We interpret the transmission spectrum for each of TOI-125 b and c through atmospheric retrievals. We use a custom code described in \cite{Espinoza:2025}, which combines the forward modelling capabilities of \texttt{POSEIDON} \citep{POSEIDON_MacDonald2023} with the \texttt{dynamic-dynesty} Nested Sampling algorithm \citep{Speagle2020_dynesty, Koposov2025_dynesty_v3.0.0}. This iteratively samples a set of atmospheric parameter values within a range of priors to generate forward models and calculate a Bayesian log-evidence value, log$Z$, for the fit of these models to the data.

We try two different retrieval assumptions: a bulk H$_2$/He-atmosphere with (1) free-chemistry, setting a log-uniform mixing ratio prior for trace molecules and (2) a chemical equilibrium model. We note that, while we do not necessarily expect the atmospheres of TOI-125 b and c to be in thermochemical equilibrium, such models can be helpful for estimating atmospheric conditions from spectra that are largely featureless. To test the robustness of our retrieval results, we also fit each spectrum to a non-physical flat line model which fits only for a reference radius and offset between NRS1 and NRS2.

For both atmospheric retrieval setups, we assume an isothermal pressure-temperature ($P$--$T$) profile and use a resolution $R=10,000$ forward model grid from 2.6 to 5.5 $\mu m$. We set uniform priors for the reference planet radius $R_\mathrm{p,ref}$ from 0.85 to 1.15 times the white-light light-curve radius, the log reference pressure $P_\mathrm{ref}$ from $-6$ to 2, and the temperature $T$ of the isothermal profile from 400--1600 K. We assume a grey cloud deck model following \cite{MacDonald&Madhusudhan2017_poseidon} and set uniform priors on the log cloud-top pressure $P_\mathrm{cloud}$ from $-6$ to 2, haze scale factor log($\alpha$) from $-4$ to 8, and haze scattering slope $\gamma$ from $-20$ to 2. We also set a uniform prior between $-1000$ to $1000$ ppm to account for a potential offset between detectors NRS1 and NRS2.

For the free-chemistry retrievals, we assume a bulk atmospheric composition of H$_2$/He with a He fraction of 0.17 $H_2$. Since TOI-125 b and c lie in a temperature regime for which we have very little observed sub-Neptune atmosphere data, we set log-uniform priors from $-12$ to 0.3 on the volume mixing ratios for trace species H$_2$O, CO, CH$_4$, CO$_2$, SO$_2$, SiO, H$_2$S, HCN, and NH$_3$. We include the same bulk and trace species in our chemical equilibrium retrievals, but set the C/O ratio and log[M/H] as free parameters with uniform priors from 0.2 to 1.2 and $-0.9$ to 3.9 respectively. A full summary of all priors is given in Table \ref{tab:retrieval_prior}.
 
\begin{table}[!ht]
    \begin{center}
        \caption{Atmospheric Retrieval Priors for TOI-125 b and c} 
        \label{tab:retrieval_prior}
        \begin{tabular}{lll}
            \hline
            Parameter&  Free Chemistry&  CEQ\\
            \hline
            $T_{\mathrm{eq}}$ (K)&  $U(400,1600)$&  $U(400,1600)$\\
            $R_{\mathrm{p,ref}}$ (m)&  $U(0.85, 1.15)R_p$&  $U(0.85, 1.15)R_p$\\
            log$P_{\mathrm{ref}}$ (bar)&  $U(-6, 2)$&  $U(-6, 2)$\\
            logX$_\mathrm{species}$&  log$U(-12, 0.3)$&   -\\
            C/O&  -&   $U(0.2, 1.2)$\\
            log[M/H]&  -&   $U(-0.9, 3.9)$\\
            log$\alpha$&  $U(-4, 8)$&  $U(-4, 8)$\\
            $\gamma$&  $U(-20, 2)$&  $U(-20, 2)$\\
            log$P_\mathrm{cloud}$ (bar)&  $U(-6, 2)$&  $U(-6, 2)$\\
            Instr. offset (ppm)&  $U(-1000, 1000)$&  $U(-1000, 1000)$\\
            \hline
        \end{tabular}
    \end{center}
    \footnotesize{The same retrieval priors are used for both planets. The flat line model only fits for $R_{\mathrm{p,ref}}$ and instrument offset; in this case, the prior values used are the same as for the atmospheric retrievals. Trace species included are H$_2$O, CO, CH$_4$, CO$_2$, SO$_2$, SiO, H$_2$S, HCN, and NH$_3$.}
\end{table}

\section{Results}\label{sec:results}

\subsection{The Transmission Spectra}

The transmission spectra resulting from the two different reductions are shown in Figure \ref{fig:2}. For TOI-125 b, the \texttt{transitspectroscopy} reduction obtains a median transit depth of $862\pm27$ ppm in NRS1 and $867\pm58$ ppm in NRS2, and the \texttt{Eureka!} reduction a median transit depth of $864\pm33$ ppm in NRS1 and $924\pm79$ ppm in NRS2. For TOI-125 c, the \texttt{transitspectroscopy} reduction obtains a median transit depth of $937\pm46$ ppm in NRS1 and $974\pm103$ ppm in NRS2, and the \texttt{Eureka!} reduction a median transit depth of $965\pm54$ ppm in NRS1 and $1018\pm136$ ppm in NRS2.

We find that the two reductions are largely consistent for both planets, with an absolute median difference in transit depth of 14 ppm in NRS1 and 58 ppm in NRS2 for TOI-125 b, and 16 ppm in NRS1 and 62 ppm in NRS2 for TOI-125 c. This is within the median 1-$\sigma$ precisions for both reductions as described above. The larger overall uncertainties in the spectra of TOI-125 c are likely due to the partial transit observation, which affects the degree of constraint on the centre of transit time and thus the degree of constraint on the wavelength-dependent transit depth. 

Interestingly, our original \texttt{PandExo} simulations provide a median TOI-125 b uncertainty of 13 ppm for NRS1 and 17 ppm for NRS2, and a median TOI-125 c uncertainty of 20 ppm for NRS1 and 33 ppm for NRS2. For NRS1, then, the measured precision appears to be largely equivalent to the predicted precision for both planets; however, NRS2 seems to perform $\sim2--3.5\times$ worse.  

We additionally find an offset of $\sim50$ ppm between the \texttt{transitspectroscopy} and \texttt{Eureka!} reductions longwards of 4.2 $\mu m$ for NRS2. The \texttt{Eureka!} reduction performs spectral extraction with a 3-pixel aperture; however, increasing the aperture from 2 to 3 pixels for the \texttt{transitspectroscopy} reduction does not noticeably change this offset. We note that residual offsets in NRS2 versus NRS1 have been observed to occur between different reductions \citep[see][]{Lothringer2025_WASP178b_HST_JWST}; as we are unable to determine an exact reason for this offset, we perform retrievals on the transmission spectra from each reduction.

\subsection{Atmospheric Retrieval Results}

In the case of largely featureless spectra, rather than providing precise constraints on specific features, retrievals can be used to narrow down a broad range of conditions that are consistent with the observed spectrum. On the whole, we find that the level of uncertainty in the spectra of both planets makes it challenging to detect molecular features and place any strong atmospheric constraints. However, given the interesting parameter space occupied by both planets, as hot sub-Neptunes around a sun-like star for which we currently have very little atmospheric information, it is still worthwhile to consider potential findings based on the retrieval results.

\subsubsection{TOI-125 b}

\begin{figure*}
    \centering
    \includegraphics[width=\linewidth]{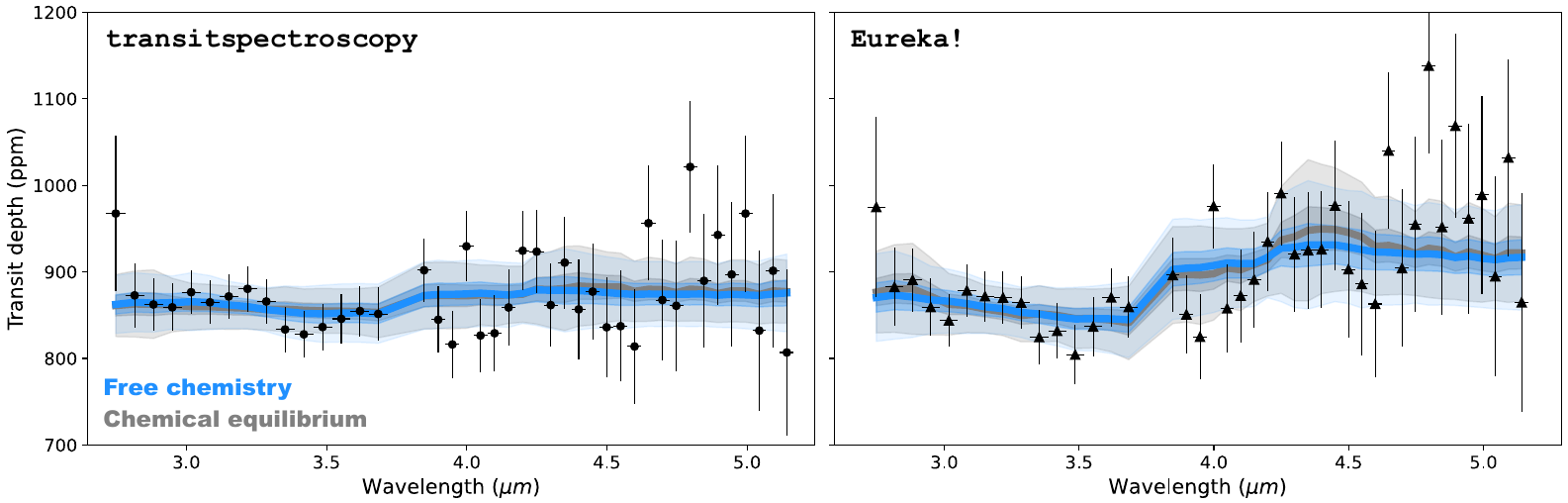}
    \caption{Best-fit retrieved models for TOI-125 b for both reductions. Free chemistry and CEQ retrieval assumptions return similar results. Lighter and darker shaded regions represent 1-$\sigma$ and 3-$\sigma$ credibility bands respectively. The larger offset between NRS1 and NRS2 in the \texttt{Eureka!} spectrum may contribute to the attempt to fit a potential 4-4.5 $\mu m$ feature; however, this only occurs in the CEQ assumption retrieval case.
    }
    \label{fig:3}
\end{figure*}

We find that the best fit models from the free-chemistry and chemical equilibrium retrievals on TOI-125 b are essentially equivalent with each other as well as a flat line ($\Delta$log$Z\sim1$ for all). The free chemistry retrieval is unable to place precise constraints on any of the included trace species, with posteriors on the VMRs spanning the entire prior range (see Appendix \ref{sec:appA} for details). We further test our results by simulating a `noise spectrum' by generating Gaussian noise based on the data uncertainties centred on the best-fit $(R_p/R_*)^2$ value from the white-light light-curve fit. After running both retrieval set-ups on the noise spectrum, we find that the posterior distributions between the data and the noise spectrum are for the most part equivalent (see Appendix \ref{sec:appB}). We do note that, in comparison to the free retrieval on noise, the data free retrieval posterior distributions for H$_2$O, CO$_2$ and NH$_3$ tend towards higher values. This may be driven by a potential downward slope/step-function from 3.0--3.5 $\mu m$ for H$_2$O and a potential broad feature from 4.0--4.5 $\mu m$ for CO$_2$. Further observations, however, are required to determine if this is truly an indication of any real molecular signals or simply a (un)fortuitous instance of noise (see Section \ref{sec:discuss}). 
In the chemical equilibrium set-up, retrievals on the data and a pure noise `spectrum' recover essentially similar posteriors (see Figure \ref{fig:3}). 

Overall, the retrievals indicate a largely featureless transmission spectrum that may be consistent with a high metallicity atmosphere and/or a high cloud deck. Using rejection contours to narrow down the region of the metallicity-cloud top pressure parameter space consistent with the observed spectra, we find that for TOI-125 b, a clear atmosphere may require a metallicity of at least 150x solar, whereas lower metallicities may require a cloud top pressure of at deepest 3 mbar (see Figure \ref{fig:5}). However, we caution that this is only compatible with the data at the 2-$\sigma$ level; placing stronger constraints on TOI-125 b's atmosphere will require more data (see Section \ref{sec:discuss}).

\subsubsection{TOI-125 c}

\begin{figure*}
    \centering
    \includegraphics[width=\linewidth]{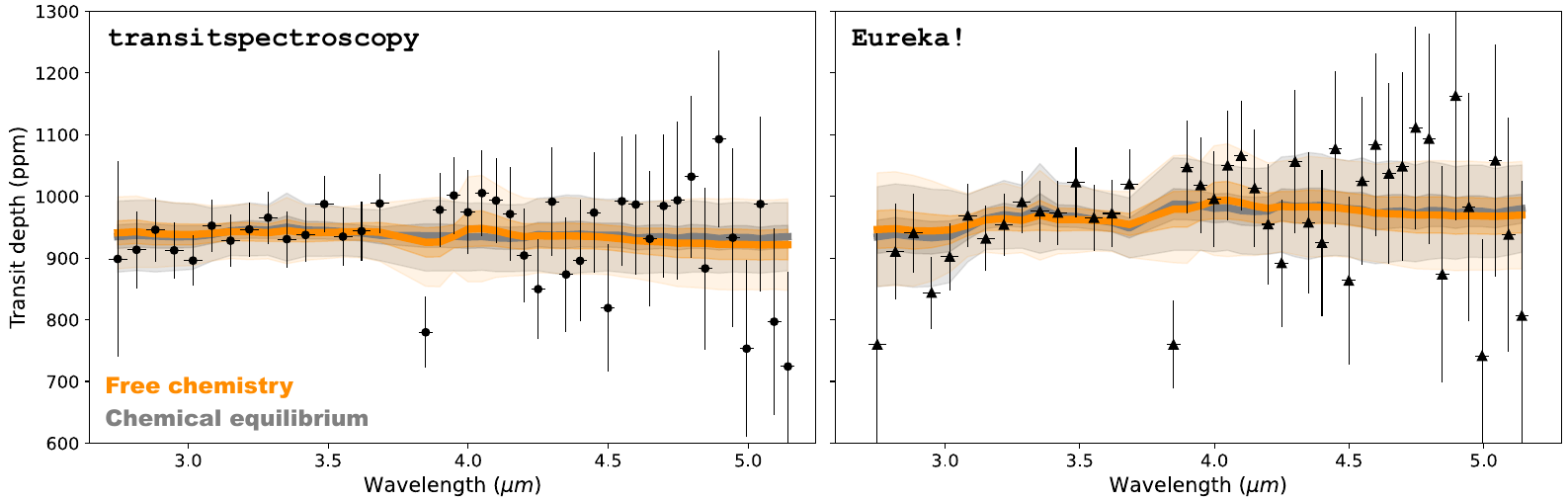}
    \caption{Best-fit retrieved models for both reductions of TOI-125 c. Free chemistry and CEQ retrieval assumptions return similar results. Lighter and darker shaded regions represent 1-$\sigma$ and 3-$\sigma$ credibility bands respectively.
    }
    \label{fig:4}
\end{figure*}

\begin{figure}
    \vspace{3mm}
    \centering
    \includegraphics[width=\linewidth]{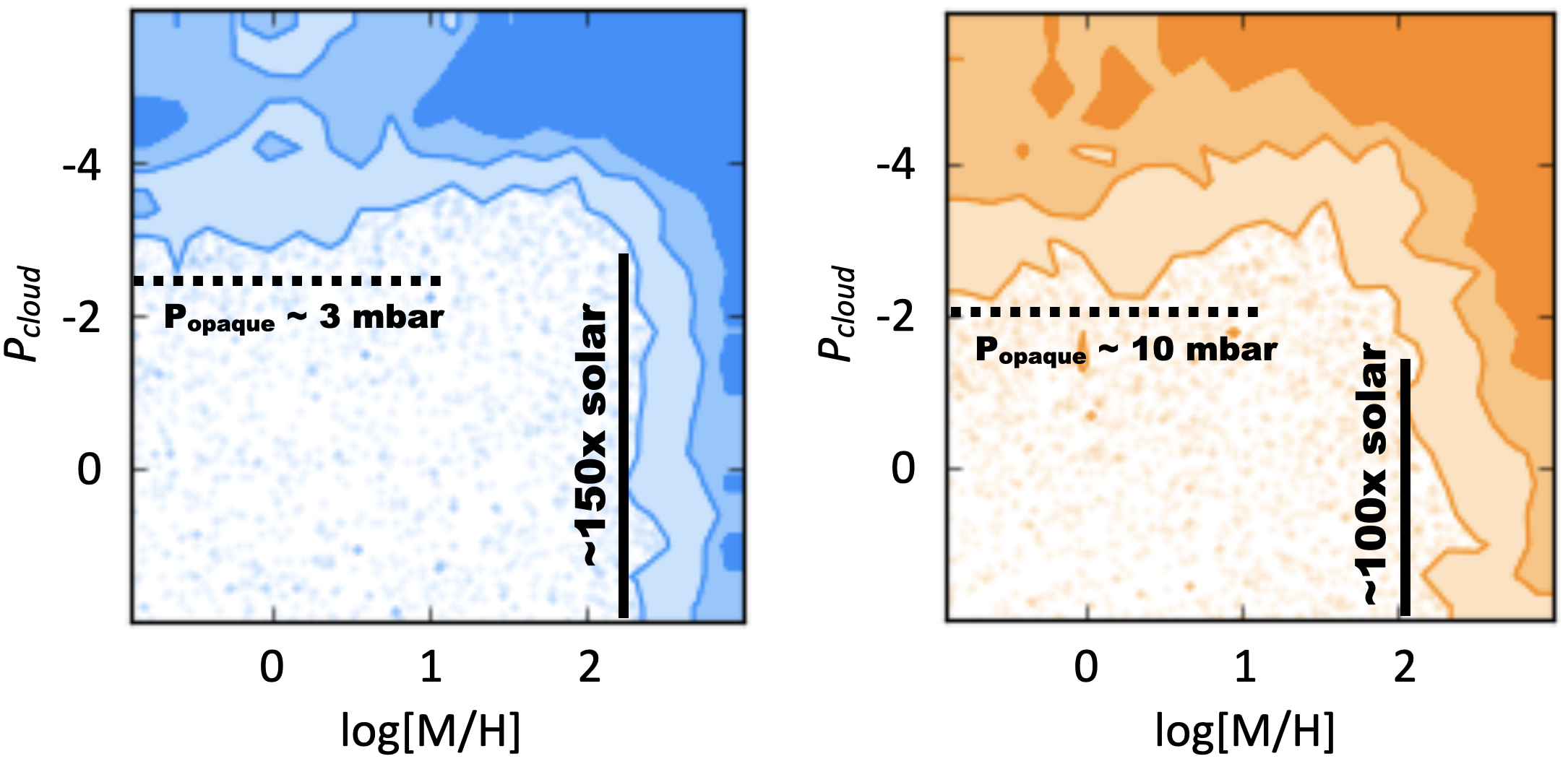}
    \caption{Rejection contours for the log[M/H]--cloud top pressure parameter space for TOI-125 b (left) and c (right), based on free chemistry retrievals for the \texttt{transitspectroscopy} reduction spectrum of each planet. Colours from darkest to lightest indicating 1-$\sigma$, 1.5-$\sigma$ and 2-$\sigma$ two-dimensional contours corresponding to 39\%, 68\% and 86\% confidence intervals. Horizontal dotted lines mark the deepest allowable pressure for an opaque cloud deck in the case of a lower metallicity atmosphere, whereas vertical black lines indicate the minimum allowable metallicity for the case of a clear atmosphere. 
    }
    \label{fig:5}
\end{figure}

Atmospheric retrievals on TOI-125 c indicate a largely featureless transmission spectrum from 3--5 $\mu m$. This is corroborated by Bayesian comparison of retrieval results with a flat model, which find that the preference for an atmospheric versus flat model is roughly equivalent (i.e., $\Delta\mathrm{log}Z\sim1$). As with TOI-125 b, we generate a noise spectrum for TOI-125 c using the data uncertainties centred about the white-light $(R_p/R_*)^2$. We find that the posterior distributions between the data and the noise spectrum are essentially equivalent (see Appendix \ref{sec:appB}), indicating that the data is consistent with a noisy flat line.

Additionally, although $T_\mathrm{eq}$ consistently tends towards the lower bound of the prior for both retrieval set-ups, this may be due to the temperature dependency of the scale height, $H=kT/\mu g$, where $k$ is the Boltzmann constant, $T$ is the temperature, $\mu$ is the MMW and $g$ is the surface gravity of the planet. A featureless spectrum, which corresponds to a small scale height, can be achieved either by increasing the MMW of the atmosphere or decreasing the planet's equilibrium temperature. As such, we consider it unlikely that the low $T_\mathrm{eq}$ values suggested for TOI-125 c are physical.

As with TOI-125 b, a featureless spectrum could indicate a metal-enriched atmosphere and/or clouds high in the atmosphere. We find that 2-$\sigma$ rejection contours are compatible with either a clear atmosphere with metallicity above 100x solar or an atmospheric metallicity lower than 100x solar with a cloud deck located at deepest at 10 mbar (see Figure \ref{fig:5}), although we caution that this does not constitute strong evidence.

\section{Discussion}\label{sec:discuss}

TOI-125 b and c add to a growing number of sub-Neptunes observed with \textit{JWST} which show largely muted or featureless 3--5 $\mu m$ transmission spectra \citep[e.g.,][]{Wallack2024_COMPASS_TOI836c, Schlawin2024a_GJ1214b_NIRSpec, Teske2025_COMPASS_TOI776c, Wallack2026AJ_COMPASS_HD15337c}. In the following discussion, we consider the implications of our transmission spectra in relation to the broader population of sub-Neptunes, along with potential avenues for follow-up observations.

\subsection{Implications for Sub-Neptune Atmosphere Trends}

Atmospheric feature sizes --- in particular, comparison of the scale-height normalised amplitude $A_H$ at a given wavelength --- have emerged as a useful metric for probing sub-Neptune trends in MMW and cloudiness/metallicity. For M-dwarf sub-Neptunes observed with \textit{HST}, \cite{Brande2024} find evidence for a parabolic trend in 1.4 $\mu m$ feature size versus equilibrium temperature, in which planets with $T_\mathrm{eq}<500$ K have feature-rich spectra, those with $500 < T_\mathrm{eq} < 850$ K have muted spectra, and those with $T_\mathrm{eq} > 850$ K once again have clear spectra. Considering higher temperatures, \cite{Ashtari2025_clearskycorridor} suggest a potential `clear sky corridor' for sub-Neptunes between 850--1000 K based on analysis of the 1.4 $\mu m$ spectral amplitude in a sample of 8 sub-Neptune and Neptune-sized planets. From aerosol theory, planets cooler than this `corridor' should show suppressed spectral features as their atmospheres begin to contain increasing amounts of methane, a driver of photochemical haze formation \citep{gao2021}. This reverses at even cooler temperatures below 500 K, where rainout allows for atmospheres to once again clear up. At the hotter edge of the corridor, atmospheres above $\sim$950 K are thought to become suppressed due to silicate condensation, although the extent of suppression may decrease with temperature as silicate clouds begin to sink below atmospheric pressure levels probed by observations \citep{Gao2020_silicatesabove950_hydrocarbonsbelow}. 

The feature-rich 1--5 $\mu m$ NIRISS SOSS $+$ NIRSpec G395M spectrum of TOI-421 b (920 K) reported by \cite{Davenport2025} seems to corroborate evidence for this clear sky corridor, conforming to expectations of a haze-free, low-MMW atmosphere. The featureless \textit{HST} WFC3 $+$ STIS 1.1--1.7 $\mu m$ spectrum of HD 86266c (1300 K) reported by \cite{Kahle2025} disrupts the proposed trend from \cite{Brande2024}, but may indicate that sub-Neptune atmospheres at even hotter temperature regimes are dominated by different process, e.g. the formation of silicate clouds as opposed to hydrocarbon aerosols.

\begin{figure}
    \centering
    \includegraphics[width=\linewidth]{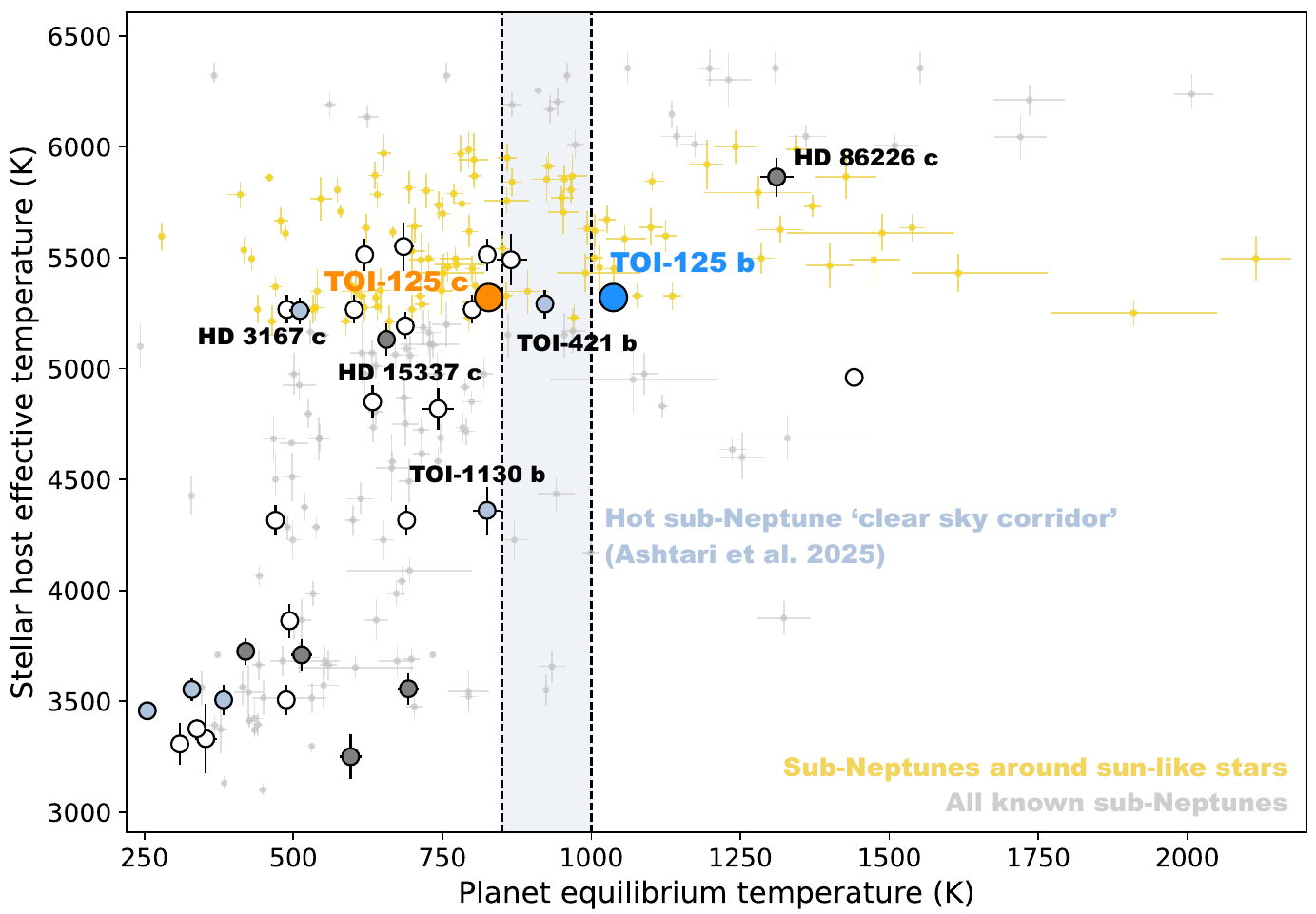}
    \caption{The population of all confirmed sub-Neptunes to date as a function of planet $T_\mathrm{eq}$ versus host star $T_\mathrm{eff}$. Light blue circles indicate planets with feature-rich spectra, grey circles indicate muted/featureless spectra, and empty circles indicate planned or unpublished \textit{JWST} observations \citep{Nikolov2022_Trexolists}. All sub-Neptunes around sun-like stars with published \textit{HST} or \textit{JWST} spectra are labelled, plus TOI-1130 b.  The $T_\mathrm{eq}$ range of the proposed `clear sky corridor' for hot sub-Neptunes from \cite{Ashtari2025_clearskycorridor} is shown in light blue; the lower bound indicates the temperature below which hydrocarbon hazes are expected to suppress spectral features, whereas the upper bound indicates the temperature above which features are obscured due to the onset of silicate cloud formation. Notably, TOI-125 b and c straddle both sides of the corridor.}
    \label{fig:subnep_pops}
\end{figure}

TOI-125 b and c, then, may lie at two very interesting crossroads in the sub-Neptune temperature parameter space (see Figure \ref{fig:subnep_pops}. At $\sim$1100 K, TOI-125 b may be located around the temperature regime at which silicate clouds may or may not be observable in the upper atmosphere. If the spectrum of TOI-125 b is indeed muted, this may provide observational evidence for the lower temperature bounds on the onset of silicate cloud formation in sub-Neptune atmospheres. Likewise, at $\sim$825 K, TOI-125 c lies just below the theoretical cutoff for efficient organic haze production due to the presence of atmospheric methane. As such, a more muted spectrum caused by hazes may still align with expectations based on the existing proposed trend. However, further data are needed before drawing any strong conclusions about the atmospheric conditions of either planet.

\subsection{Observing Mode Limitations and Potential for Follow-Up}

The NIRSpec/G395H observing mode has been used by many \textit{JWST} exoplanet atmosphere programs due to its coverage of key carbon-, oxygen- and sulphur-bearing species, as well as its higher level of precision that is key to recovering the smaller relative feature sizes of smaller planets. We find that our limited ability to detect active atmospheric molecules with the NIRSpec/G395H 3--5 $\mu m$ data alone echoes the findings of \cite{Davenport2025}. As such, we emphasise the need for shorter wavelength observations before drawing definite conclusions about the atmospheres of TOI-125 b and TOI-125 c.

\begin{figure}[!th]
    \centering
    \includegraphics[width=\linewidth]{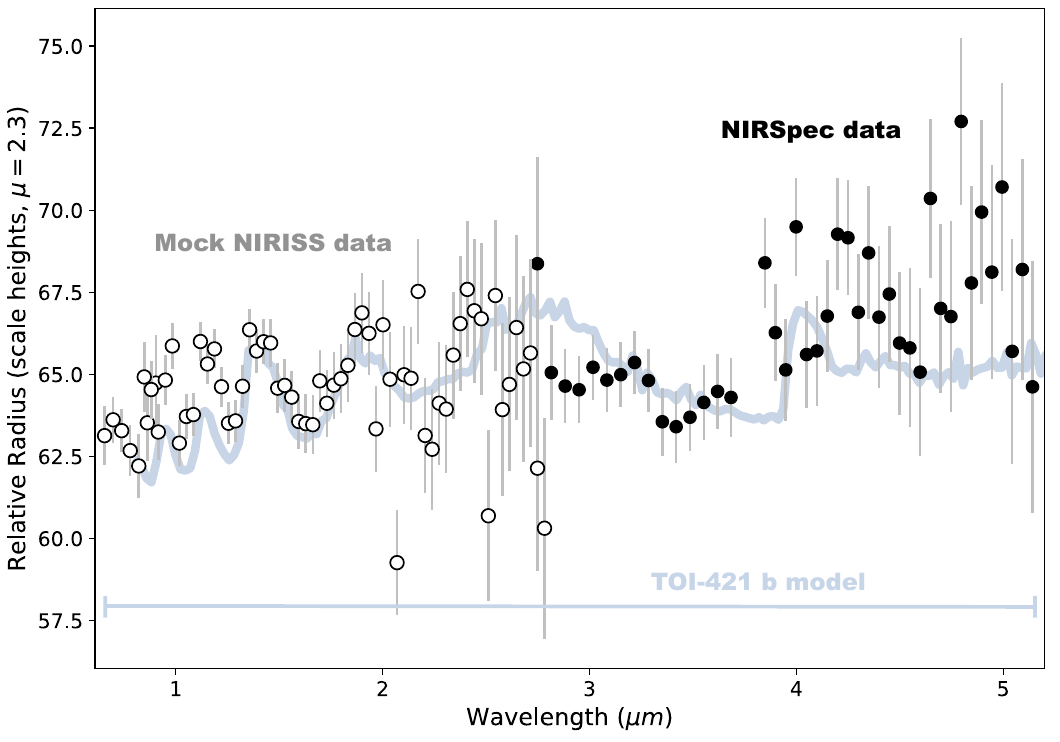}
    \caption{Scale-height normalised \texttt{PandExo} simulation of a feature-rich NIRISS SOSS spectrum based on a single transit observation of TOI-125 b. The existing NIRSpec G395H spectrum from the\texttt{transitspectroscopy} reduction and the best-fit model spectrum for TOI-421 b from \cite{Davenport2025} are shown for comparison. 
    }
    \label{fig:followup_obs}
\end{figure}

To demonstrate the impact of shorter wavelength observations, we use \texttt{PandExo} \citep{batalha2017_pandexo} to simulate NIRISS SOSS 1--3 $\mu m$ data for TOI-125 b for two cases: (1) a flat model and (2) a clear, solar metallicity atmosphere consistent with TOI-421 b. Figure \ref{fig:followup_obs} shows a comparison of the existing NIRSpec and mock feature-rich NIRISS SOSS data for TOI-125 b to the best-fit atmospheric model for TOI-421 b. Currently, with the NIRSpec data alone, retrievals are unable to strongly distinguish between an atmosphere and a flat model. However, with the addition of NIRISS SOSS data from a single transit observation, retrievals overwhelmingly prefer the correct input model, with Bayesian log-evidence ratios $B\sim10^7$ for the atmospheric model case and $B
\sim300$ for the flat model case.

In addition to the high potential information content provided by NIRISS SOSS follow-up, the value of shorter wavelength observations becomes further apparent when considering the unique position both planets occupy in the sub-Neptune parameter space. As previously discussed, we have only just begun to probe the atmospheres of hot sub-Neptunes around sun-like stars. The three data points currently pubished --- TOI-421 b, HD 86226 c and TOI-1130 b --- have already hinted that the atmospheres of sub-Neptunes in this temperature regime may be different from those of relatively cooler sub-Neptunes around M- and K-dwarfs. 

1--3 $\mu m$ observations of TOI-125 b, and perhaps also c, could offer a prime opportunity to further our understanding of hot sub-Neptune atmospheres. By determining whether the spectra of these two planets are feature-rich, like TOI-421 b and TOI-1130 b, or truly feature-poor, like HD 86226 c, we may be able to place observational constraints on the proposed `clear sky corridor' for sub-Neptunes. We note that, should TOI-125 b display featureless spectra at NIRISS SOSS wavelengths, mid-IR observations targeting the silicate absorption bandhead at 8 $\mu m$, with e.g. MIRI LRS, may help to determine whether the muting of spectra is due to the onset of silicate cloud formation at temperatures $>1100$ K, as suggested in \cite{Kahle2025}.

Finally, compared to other potential hot sub-Neptune targets, the TOI-125 system offers the benefit of being relatively bright ($m_J=9.466\pm0.021$); following Equation 1 of \cite{Kempton2018_TSM_ESM}, this gives TOI-125 b and c transmission spectroscopy metrics of 51 and 64 respectively. TOI-125 is also one of few systems with multiple confirmed sub-Neptunes orbiting a sun-like host --- and the only such system to have spectroscopic measurements to date. This shared location within a single system controls for formation within the same protoplanetary disk, enabling truly comparative study. Thus, we consider both TOI-125 b and c to be prime targets for follow-up observations, particularly TOI-125 b at shorter wavelengths with \textit{JWST} NIRISS SOSS. 

\section{Conclusions}\label{sec:concl}

In this work, we report the first spectroscopic observations of TOI-125 b and c, two hot sub-Neptunes within a single system orbiting a sun-like star. Both planets lie in a regime of the sub-Neptune parameter space for which there is currently very little atmospheric knowledge. To date, there are only 2 other hot ($T_\mathrm{eq}>800 K$) sub-Neptunes around sun-like stars with published spectra. These are already beginning to hint at a diversity of atmospheric properties that challenge previously proposed trends in sub-Neptune feature size with temperature for cooler planets around M- and K-dwarfs. This highlights the need for additional data to better understand the nature of this intriguing class of planets.

To this end, we observe one transit of each planet with \textit{JWST} NIRSpec G395H in late 2024. We obtain the full transit of TOI-125 b, but only 70\% of the transit of TOI-125 c due to larger than predicted TTVs. To investigate the orbital parameters of the system, we perform a global fit using light curves from \textit{JWST} along with those from TESS sectors 2, 28, 29, 68 and 69 and HARPS RV data. We fit a linear trend to the TTVs and find variations on the order of 1.5 hours for TOI-125 b, 2.5 hours for TOI-125 c, and 1 hour for TOI-125 d. Although presently outside the scope of this work, photodynamical modelling may be useful for investigating the dynamics of the system in greater detail, including determining whether the larger TTVs indicate the presence of an additional unconfirmed planet(s) within the system as suggested by \cite{Quinn2019_TOI125_discovery}. 

We then present the first 3--5 $\mu m$ $R\sim100$ transmission spectra for both planets, which both appear largely featureless, adding to a growing sample of muted/featureless sub-Neptunes at these wavelengths. Our atmospheric retrievals may be consistent with a $\sim$150x solar metallicity atmosphere or clouds above $\sim$3 mbar for TOI-125 b, and a $\sim$100x solar metallicity atmosphere or clouds above $\sim$10 mbar for TOI-125 c. 

TOI-125 b and c are uniquely placed to provide information on sub-Neptunes at hotter temperature regimes. At $T_\mathrm{eq}\sim1100$K and $\sim825$ K respectively, they straddle the predicted bounds of spectral obscuration due to hydrocarbon haze production ($<850$ K) and silicate cloud formation ($>1000$ K). If both planets do possess featureless spectra, this could help place observational constraints on the upper and lower temperature bounds of the proposed `clear sky corridor' for sub-Neptune atmospheres. For TOI-125 b in particular, additional data at shorter wavelengths (e.g. with NIRISS SOSS) will help to determine whether its transmission spectra is truly featureless or instead feature-rich, with implications for the nature of hot sub-Neptune atmospheres around FGK stars.

\section{Software and third party data repository citations} \label{sec:cite}

All the {\it JWST} and {\it TESS} data used in this paper can be found in MAST: \dataset[10.17909/w4dh-3861]{http://dx.doi.org/10.17909/w4dh-3861}.

\begin{acknowledgments}

This work is based [in part] on observations made with the NASA/ESA/CSA James Webb Space Telescope. The data were obtained from the Mikulski Archive for Space Telescopes at the Space Telescope Science Institute, which is operated by the Association of Universities for Research in Astronomy, Inc., under NASA contract NAS5-03127 for JWST. These observations are associated with program \#4126.

Support for program \#4126 was provided by NASA through a grant from the Space Telescope Science Institute, which is operated by the Association of Universities for Research in Astronomy, Inc., under NASA contract NAS5-03127.

EMV acknowledges financial support from the Swiss National Science Foundation (SNSF) Postdoctoral Mobility Fellowship under grant number P500PT\_225456/1.

B.-O. D. acknowledges support from the Swiss State Secretariat for Education, Research and Innovation (SERI) under contract number MB22.00046.

KH is partially supported by the European Research Council (ERC) Synergy Grant Geoastronomy (grant number: 101166936).

MLM is supported by individual research time under NASA contracts NAS5-26555 and NAS5-03127 to the Associated Universities for Research in Astronomy for the operation of the Hubble Space Telescope and the James Webb Telescope Science Operations Centers at STScI.

We also acknowledge the co-investigators of the original observing proposal for Cycle 2 program \#4126, which led to the \textit{JWST} data used in this work: Hannah Diamond-Lowe, Jens Hoeijmakers, Andrea Guzman Mesa, Daniel Kitzmann, Neale Gibson, Lars A. Buchhave, Joao Manuel Mendonca, Alexander Rathcke, Thea Kozakis, Yann Alibert, Mark Fortune, Kathryn Jones, Bibiana Prinoth, Nicholas Borsato, Anna Lueber, and Can Jan Akin.

Finally, we thank the anonymous referee for their helpful feedback which improved this work.

\end{acknowledgments}

\begin{contribution}

YC was responsible for leading the writing and submission of the manuscript. NE provided feedback both throughout the analysis and on the final manuscript, including text edits and suggestions for figures. CF provided the observations which gave rise to this work and contributed to the atmospheric retrievals and analysis. EMV led an independent reduction of the data and provided feedback on the manuscript. MH ran a preliminary reduction of the data. BM provided feedback throughout the analysis. NHA, BOD, AG, KH, MLM and MT provided feedback on the manuscript.

\end{contribution}

%
\facilities{ \textit{JWST}(NIRSpec), TESS, ESO(HARPS)} 

\software{\textit{JWST} Calibration Pipeline \citep{Bushouse2025_jwst_pipeline_1.17.1}, transitspectroscopy \citep{transitspectroscopy}, Eureka! \citep{Bell2022_Eureka}, juliet \citep{Espinoza2019_juliet}, ExoTiC-LD \citep{Grant&Wakeford2024_exotic-ldc}, POSEIDON \citep{POSEIDON_MacDonald2023}, dynesty \citep{Speagle2020_dynesty, Koposov2025_dynesty_v3.0.0}, PandExo \cite{batalha2017_pandexo}
}

\appendix

\section{TOI-125 System Global Fit and TTVs}\label{sec:appA}

Since the initial TESS discovery and follow-up mass measurements of the TOI-125 system reported by \cite{Quinn2019_TOI125_discovery} and \cite{Nielson2020_TOI125_masses}, it has been observed in an additional 3 TESS sectors. We therefore perform a global fit on all existing data for the TOI-125 system in order to obtain updated parameters for all three confirmed planets. We use the \texttt{juliet} library (version 2.2.8) to simultaneously fit the light curves from TESS sectors 1, 2, 28, 68, 69, the \textit{JWST} NIRSpec G395H NRS1 and NRS2 light curves for planets b and c, and the HARPS radial velocities. We fit for eccentricity, $e$, and argument of periastron, $\omega$, using the parametrisations $\sqrt{e}\mathrm{cos}\omega$ and $\sqrt{e}\mathrm{sin}\omega$. To account for TTVs, we directly fit for the time of transit, $T(n)=t_0+nP+\delta t_n$, where $t_0$ is the centre of transit time, $P$ is the period of the orbit, $n$ is the transit epoch and $\delta t_n$ is the variation in transit time due to gravitational perturbation from additional bodies in the system (e.g. other planets). For each transit of TOI-125 b, c and d across all light curve data, we set a uniform prior $\pm 0.03$ BJD around an initial guess on the centre of transit time for that particular transit. We also include a fit to a linear model to account for non-physical systematics in the data. Thus the total signal modelled by \texttt{juliet} is described by

\begin{equation}
    M_i(t)+\mathrm{LM}_i(t) + \epsilon_i(t)
\end{equation}

where $M_i(t)$ is the model for the deterministic transit or radial-velocity signals, $\epsilon_i(t)$ is the model for noise, and $\mathrm{LM}_i(t)$ is the linear model defined as

\begin{equation}
    \mathrm{LM}_i(t) = \sum^{p_i}_{n=0} x_{n,i}(t)\theta^\mathrm{LM}_{n,i}
\end{equation}

Here, $x_{n,i}(t)\theta$ are the linear regressors at time $t$ for instrument $i$, and $\theta^\mathrm{LM}_{n,i}$ are their coefficients.

\begin{figure}[!ht]
    \centering
    \includegraphics[width=\linewidth]{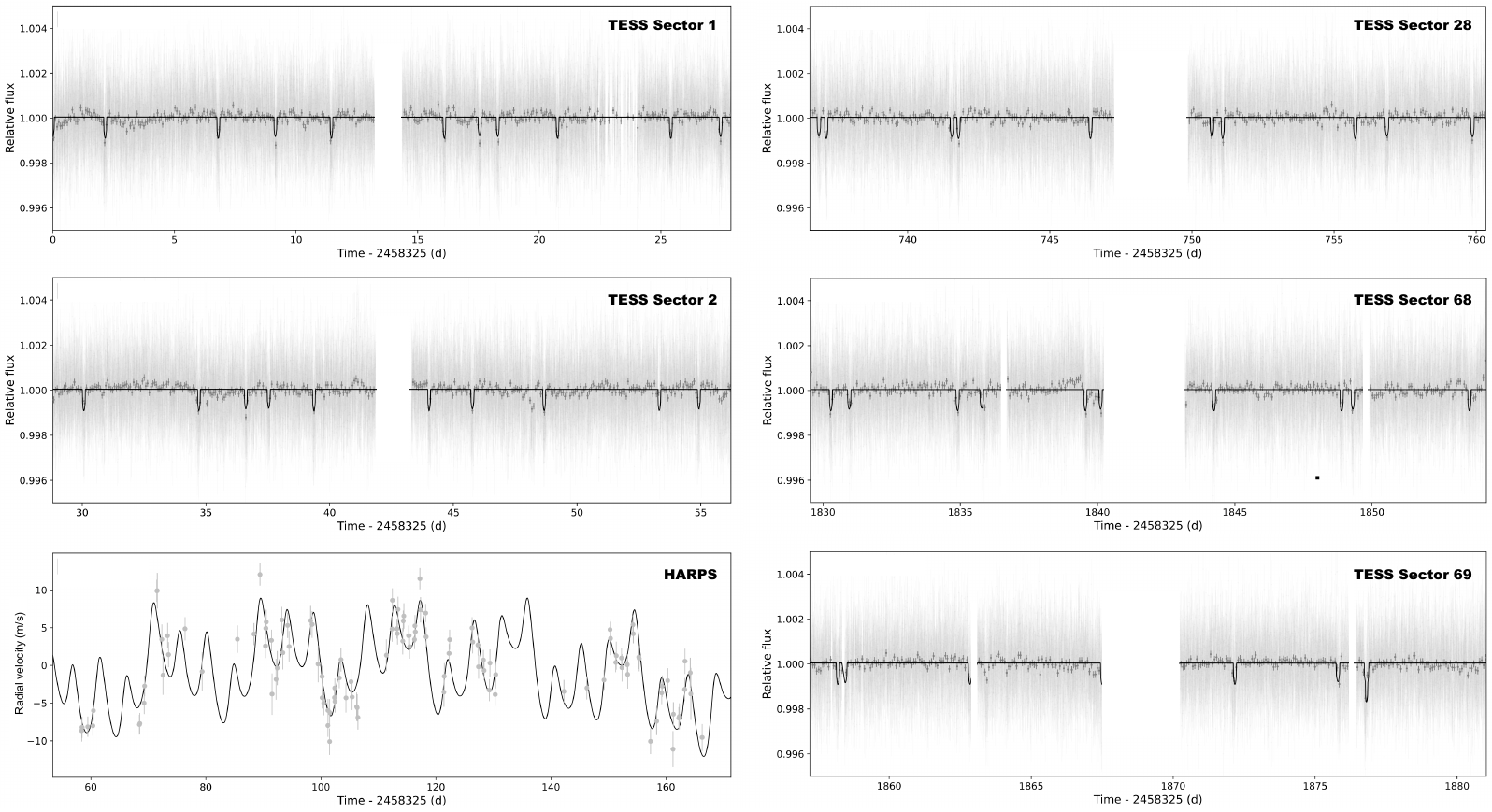}
    \caption{Global fit for the TOI-125 system using light curves from TESS Sectors 1, 2, 28, 68, 69 and JWST/NIRSpec along with HARPS radial velocities. Data is shown in light grey; dark grey points (bin size $=$ 50) for the TESS light curves are shown for ease of comparison to the model shown in black.}
    \label{fig:globalfit}
\end{figure}

Table \ref{tab:globalfit} summarises the priors used in the global fit, and the resulting posterior or derived parameters for the TOI-125 system. $P$, $t_0$, $e$, and $\omega$ are output by \texttt{juliet}. All other derived parameters are calculated using the median posteriors of the relevant parameters, or literature values in the case of $R_*$, plus propagated uncertainties. Equilibrium temperatures $T_\mathrm{eq}=T_*\sqrt{R_*/2a}$ assume negligent albedo. Inclination is calculated accounting for eccenticity using $\cos i=b\left(\frac{R_*}{a}\right)\left(\frac{1+e\sin\omega}{1-e^2}\right)$. Planet mass is $M_p\sin{i}\approx M_*^{2/3}\left(\frac{PK^3}{2\pi G}\right)^{1/3}$, where $G$ is the gravitational constant.  The results of our global fit are shown in Figure \ref{fig:globalfit} (excepting the fit to the white light JWST/NIRSpec light curves, which are shown in Figure \ref{fig:1}). 

\begin{table*}[!ht]
    \begin{center}
        \caption{Global Fit Prior and Posterior Parameters for the TOI-125 System} 
        \label{tab:globalfit}
        \begin{tabular}{lllll}
            \hline
            Planet Parameters& Prior& TOI-125 b&  TOI-125 c& TOI-125 d\\
            \hline
            $T(n)$ (BJD)& $U(T_\mathrm{n}-0.03,T_\mathrm{n}+0.03)$& *& *& *\\
            $P$ (d)& -& $4.651734^{+0.000004}_{-0.000004}$& $9.15498^{+0.00001}_{-0.00001}$& $19.98087^{+0.00003}_{-0.00003}$\\
            $t_0$ (BJD)& -& $2458327.4507^{+0.0009}_{-0.0009}$& $2458325.291^{+0.002}_{-0.002}$& $2458342.854^{+0.002}_{-0.003}$\\ 
            $b$& $U(0, 1)$& $0.58^{+0.06}_{-0.06}$& $0.62^{+0.08}_{-0.09}$& $0.47^{+0.15}_{-0.21}$\\
            $\sqrt{e}\mathrm{cos}\omega$& $U(-1, 1)$& $0.37^{+0.04}_{-0.04}$& $-0.02^{+0.14}_{-0.14}$& $0.27^{+0.07}_{-0.09}$\\
            $\sqrt{e}\mathrm{sin}\omega$& $U(-1, 1)$& $-0.24^{+0.10}_{-0.08}$& $-0.16^{+0.21}_{0.27}$& $0.14^{+0.20}_{-0.23}$\\
            $e$& -& $0.20^{+0.03}_{-0.03}$& $0.07^{+0.13}_{-0.05}$& $0.12^{+0.05}_{-0.04}$\\
            $\omega$ (deg)& -& $-33^{+14}_{-11}$& $-82^{+165}_{-37}$& $26^{+34}_{-42}$\\
            $a/R_*$& $U(1, 100)$& $11.7^{+0.78}_{-0.76}$& $20.8^{+2.46}_{-2.23}$& $42.1^{+5.56}_{-5.66}$\\
            $a$ (au)& -& $0.046^{+0.003}_{-0.003}$& $0.0820^{+0.010}_{-0.009}$& $0.166^{+0.022}_{-0.022}$\\
            $R_p/R_*$ TESS& $U(0.02, 0.04)$& $0.0291^{+0.0007}_{-0.0007}$& $0.0283^{+0.0010}_{-0.0009}$& $0.0265^{+0.0015}_{-0.0013}$\\
            $R_p$ TESS ($R_\oplus$)& -& $2.69^{+0.07}_{-0.07}$& $2.62^{+0.09}_{-0.08}$& $2.45^{+0.14}_{-0.12}$\\
            $R_p/R_*$ NRS1& $U(0.02, 0.04)$& $0.0292^{+0.0003}_{-0.0003}$& $0.0308^{+0.0004}_{-0.0004}$& -\\        
            $R_p$ NRS1 ($R_\oplus$)& -& $2.70^{+0.03}_{-0.03}$& $2.85^{+0.04}_{-0.04}$& -\\        
            $R_p/R_*$ NRS2& $U(0.02, 0.04)$& $0.0297^{+0.0005}_{-0.0005}$& $0.0308^{+0.0007}_{-0.0007}$& -\\
            $R_p$ NRS2 ($R_\oplus$)& -& $2.74^{+0.05}_{-0.05}$& $2.85^{+0.06}_{-0.06}$& -\\
            $K$ (m/s)& $U(0.0, 100.0)$& $4.33^{+0.34}_{-0.30}$& $2.31^{+0.25}_{-0.27}$& $3.64^{+0.24}_{-0.25}$\\
            $T_\mathrm{eq}$ (K)& -& $1100^{+35}_{-38}$& $825^{+45}_{-48}$& $580^{+35}_{-43}$\\
            $i$ (deg)& -& $87.4^{+0.3}_{-0.3}$& $88.4^{+0.3}_{-0.3}$& $89.4^{+0.2}_{-0.3}$\\
            $M_p$ ($M_\oplus$)& -& $10.0^{+0.9}_{-0.8}$& $6.8^{+0.8}_{-0.8}$& $13.9^{+1.0}_{-1.0}$\\
            \hline
            RV Parameters& Prior& Value&  & \\
            \hline
            RV intercept& Fixed& 0& & \\
            RV slope& $U(-100, 100)$& $-0.0954^{+0.0101}_{-0.0101}$&  & \\
            RV quadratic term& $U(-100, 100)$& $-0.00190^{+0.00022}_{-0.00022}$&  & \\
            \hline
            Instrumental Parameters& Prior& Value&  & \\
            \hline  
            $u_1$ TESS& $U(-3, 3)$& $1.09^{+0.316}_{-0.357}$&  & \\
            $u_2$ TESS& $U(-3, 3)$& $-0.539^{+0.465}_{-0.425}$&  & \\
            $u_1$ NRS1& $U(-3, 3)$& $-0.232^{+0.328}_{-0.356}$&  & \\
            $u_2$ NRS1& $U(-3, 3)$& $0.497^{+0.421}_{-0.400}$&  & \\
            $u_1$ NRS2& $U(-3, 3)$& $-0.141^{+0.484}_{-0.558}$&  & \\
            $u_2$ NRS2& $U(-3, 3)$& $0.374^{+0.641}_{-0.559}$&  & \\
            $m_\mathrm{dilution}$ TESS& Fixed& 1.0&  & \\
            $m_\mathrm{dilution}$ NRS1& Fixed& 1.0&  & \\
            $m_\mathrm{dilution}$ NRS2& Fixed& 1.0&  & \\
            $M_\mathrm{flux}$ TESS& $N(0.0, 0.1)$& $-0.0000415^{+0.00000306}_{-0.00000310}$&  & \\
            $M_\mathrm{flux}$ NRS1& $N(0.0, 0.1)$& $-0.000256^{+0.00000663}_{-0.00000647}$&  & \\
            $M_\mathrm{flux}$ NRS2& $N(0.0, 0.1)$& $-0.000289^{+0.0000133}_{-0.0000130}$&  & \\
            $\sigma_\omega$ TESS (ppm)& log$U(0, 10000)$& $6.09^{+14.37}_{-4.07}$&  & \\
            $\sigma_\omega$ NRS1 (ppm)& log$U(0, 10000)$& $217.04^{+4.38}_{-4.04}$&  & \\
            $\sigma_\omega$ NRS2 (ppm)& log$U(0, 10000)$& $459.94^{+7.63}_{-7.45}$&  & \\
            $\sigma_\omega$ HARPS (m/s)& log$U(0.001, 100.0)$& $1.70^{+0.17}_{-0.17}$&  & \\
            $\mu$ HARPS (m/s)& $U(-100, 100)$& $0.128^{+0.245}_{-0.252}$&  & \\
            \hline
            Linear Model Parameters& Prior& Value&  & \\
            \hline  
            $\theta_0$ TESS& $U(-10, 10)$& $-0.000000727^{+0.00000379}_{-0.00000381}$&  & \\
            $\theta_0$ NRS1& $U(-10, 10)$& $-0.000000727^{+0.00000379}_{-0.00000381}$&  & \\
            $\theta_0$ NRS2& $U(-10, 10)$& $0.00000588^{+0.0000130}_{-0.0000129}$&  & \\
        \end{tabular}
    \end{center}
    \footnotesize{$^*$ TTVs are accounted for by fitting directly for transit times $T(n)$, using a uniform prior $\pm0.003$ BJD centred on an initial guess for $t_0$ from the data. Since we fit directly for $T(n)$, we do not set priors on period $P$ or centre of transit timing $t_0$; these are derived from the best-fit values to $T(n)$.}
    \vspace{-5mm}
\end{table*}

We find that, in general, our derived planet parameters are consistent with those previously reported in the literature by \cite{Quinn2019_TOI125_discovery} and \cite{Nielson2020_TOI125_masses}. The addition of transit epochs from TESS sectors 28, 68, 69 and the \textit{JWST} observations of TOI-125 b and c decrease uncertainties on the periods of all three planets by a factor of $\gtrsim$4. We note that, while our radius values for TOI-125 b and c from the TESS light curves are consistent with those given in \cite{Quinn2019_TOI125_discovery}, albeit somewhat smaller in size, our derived radius for TOI-125 d (2.45 $R_\oplus$) is smaller than the 2.93$\pm$0.17 $R_\oplus$ value reported by \cite{Nielson2020_TOI125_masses}.

\begin{figure}[!ht]
    \centering
    \includegraphics[width=0.8\linewidth]{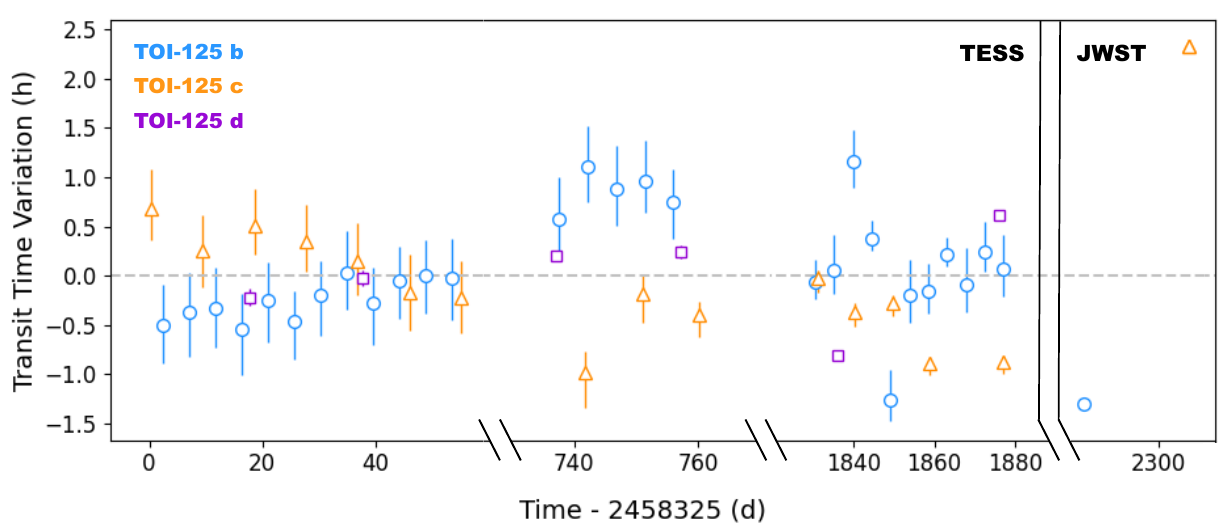}
    \caption{Updated TTV estimates for the three planets in the TOI-125 system. Blue circles indicate the variation $\delta t_n$ in hours for TOI-125 b, which ranges from 1.3 hours earlier to 1.16 hours later than the expected time of transit for a periodic orbit. Orange triangles for TOI-125 c show transit times ranging from 1.0 hour earlier to 2.33 hours later than expected. Purple squares for TOI-125 d show transit times ranging from 37 minutes earlier to 49 minutes later than expected. The data used in this figure is available.}
    \label{fig:6}
\end{figure}

We also use the $T(n)$, $P$ and $t_0$ values from the global fit to calculate updated TTVs for all three planets; we show our results for $\delta t_n$ in Figure \ref{fig:6}. It becomes apparent that TTV estimates based on only TESS sectors 1 and 2 significantly underestimate the extent of variation in $\delta t_n$ for all three planets, thus affecting our ability to accurately predict the ephemerides for TOI-125 c for our observations. We find TTVs up to $\pm$1.30 hours for TOI-125 b, $\pm$2.33 hours for TOI-125 c and $\pm$0.81 hours for TOI-125 d. The larger than anticipated TTVS may indicate the presence of a yet-undetected additional planet in the system; more detailed non-linear modelling is needed to investigate the possibility of a multi-periodic oscillation in time, but is presently outside the scope of this work.

\section{Retrieval Corner Plots}\label{sec:appB}

Here we include the full corner plots from each of our atmospheric retrieval cases to show the posterior distributions for all free parameters.

As an additional test of whether our retrievals provide any information on the data, we also use the same set-up described in Section \ref{sec:methods} to perform retrievals and a flat line fit on pure noise. We obtain a noise `spectrum' for each planet by generating Gaussian noise at each data wavelength bin, centred on the white-light light-curve $R_p/R_*$ value with standard deviations corresponding to the data uncertainties.

In each of the following corner plots, we show the retrieval posteriors for the transmission spectra from both data reductions as well as the noise spectrum. 

\begin{figure}[!h]
    \centering
    \includegraphics[width=\linewidth]{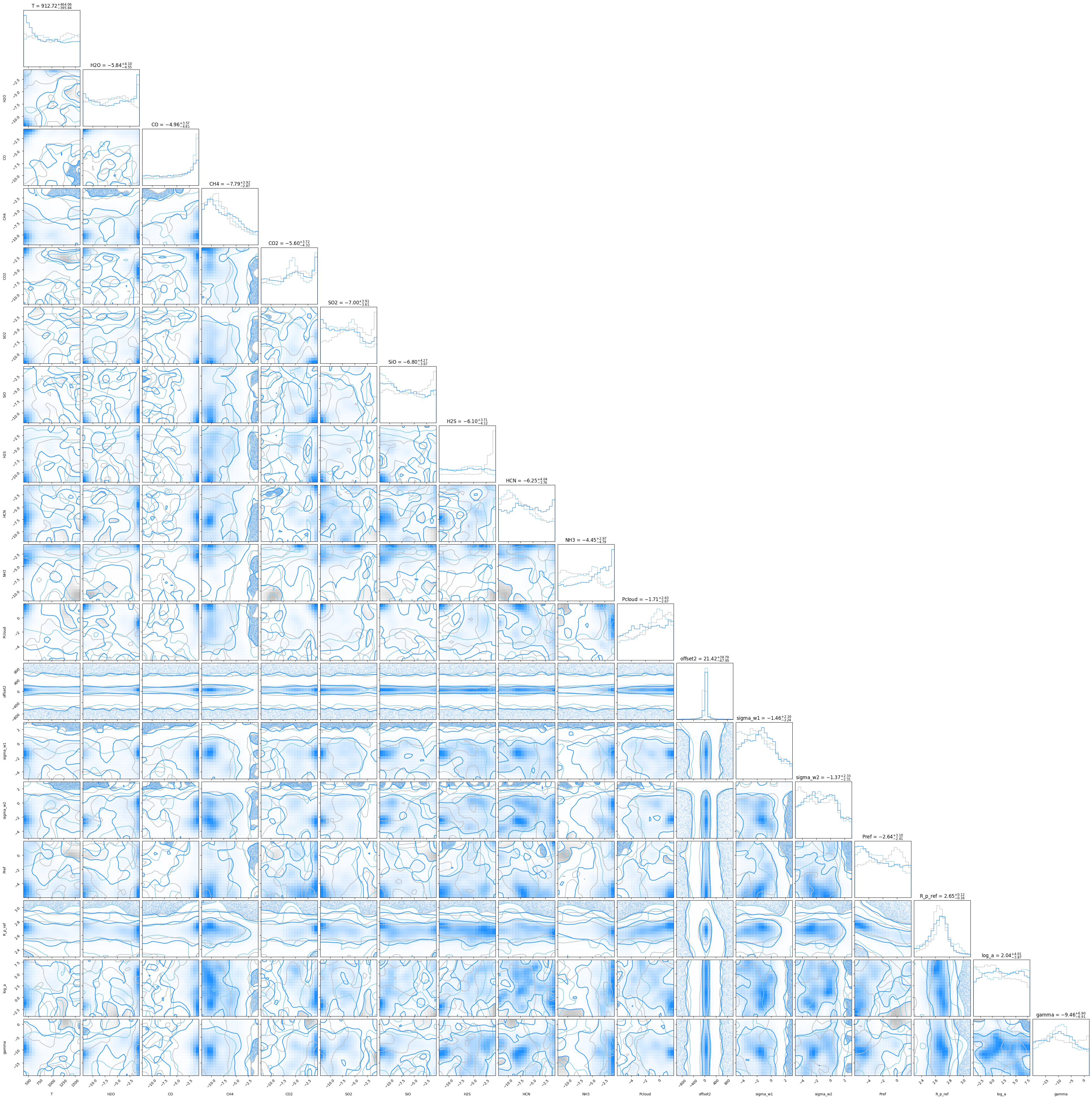}
    \caption{Isothermal, free-chemistry retrieval for TOI-125 b, assuming bulk H$_2$/He atmosphere with a log-uniform prior on the mixing ratio of all trace molecular species. Dark blue indicates posterior results from the \texttt{transitspectroscopy} reduction, light blue from the \texttt{Eureka!} reduction, and grey from the noise spectrum.}
    \label{fig:Cb1}
\end{figure}

\begin{figure}
    \centering
    \includegraphics[width=\linewidth]{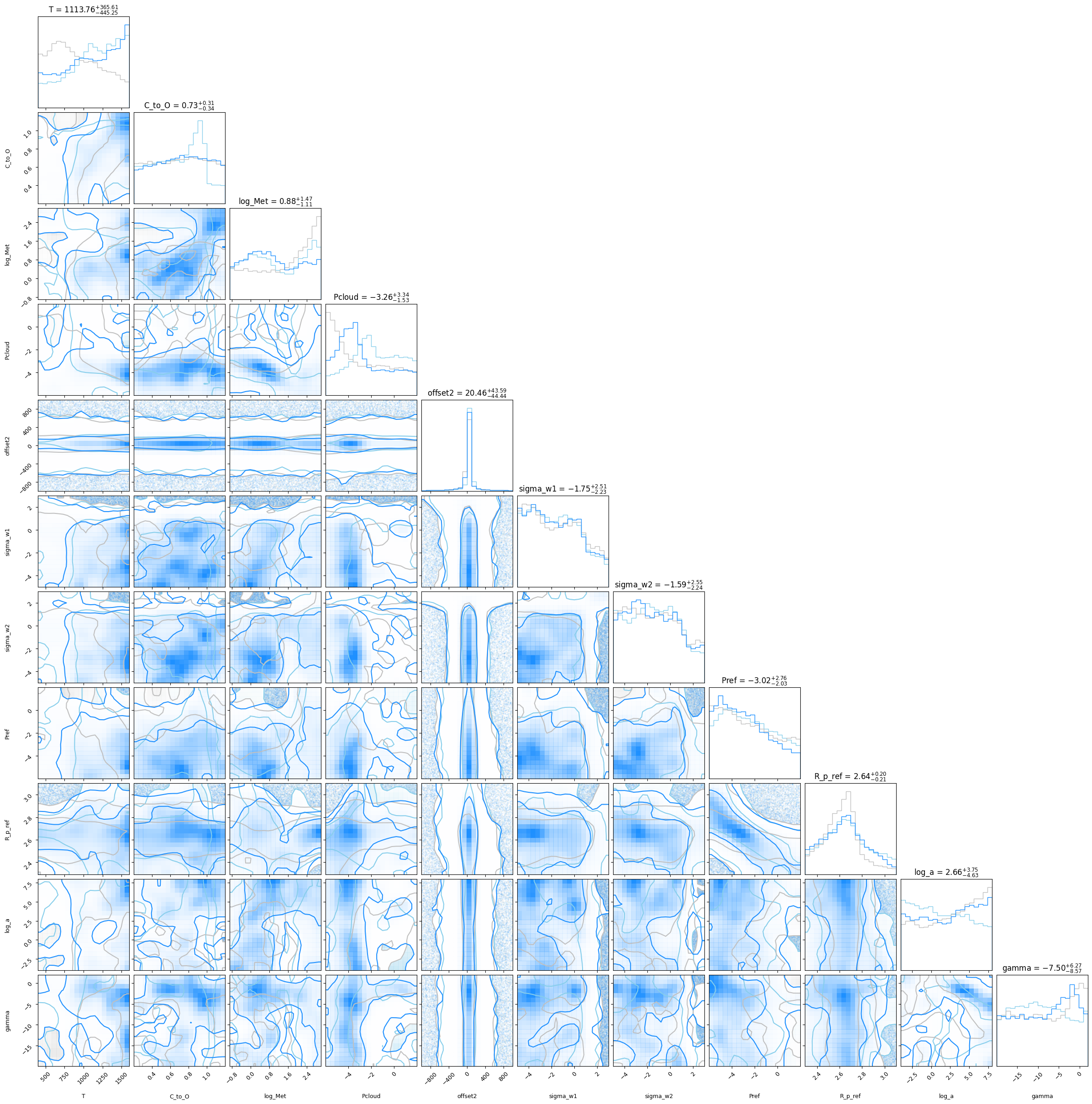}
    \caption{\texttt{POSEIDON} isothermal, chemical equilibrium retrieval for TOI-125 b. Dark blue indicates posterior results from the \texttt{transitspectroscopy} reduction, light blue from the \texttt{Eureka!} reduction, and grey from the noise spectrum. 
    }
    \label{fig:Cb3}
\end{figure}

\begin{figure}
    \centering
    \includegraphics[width=\linewidth]{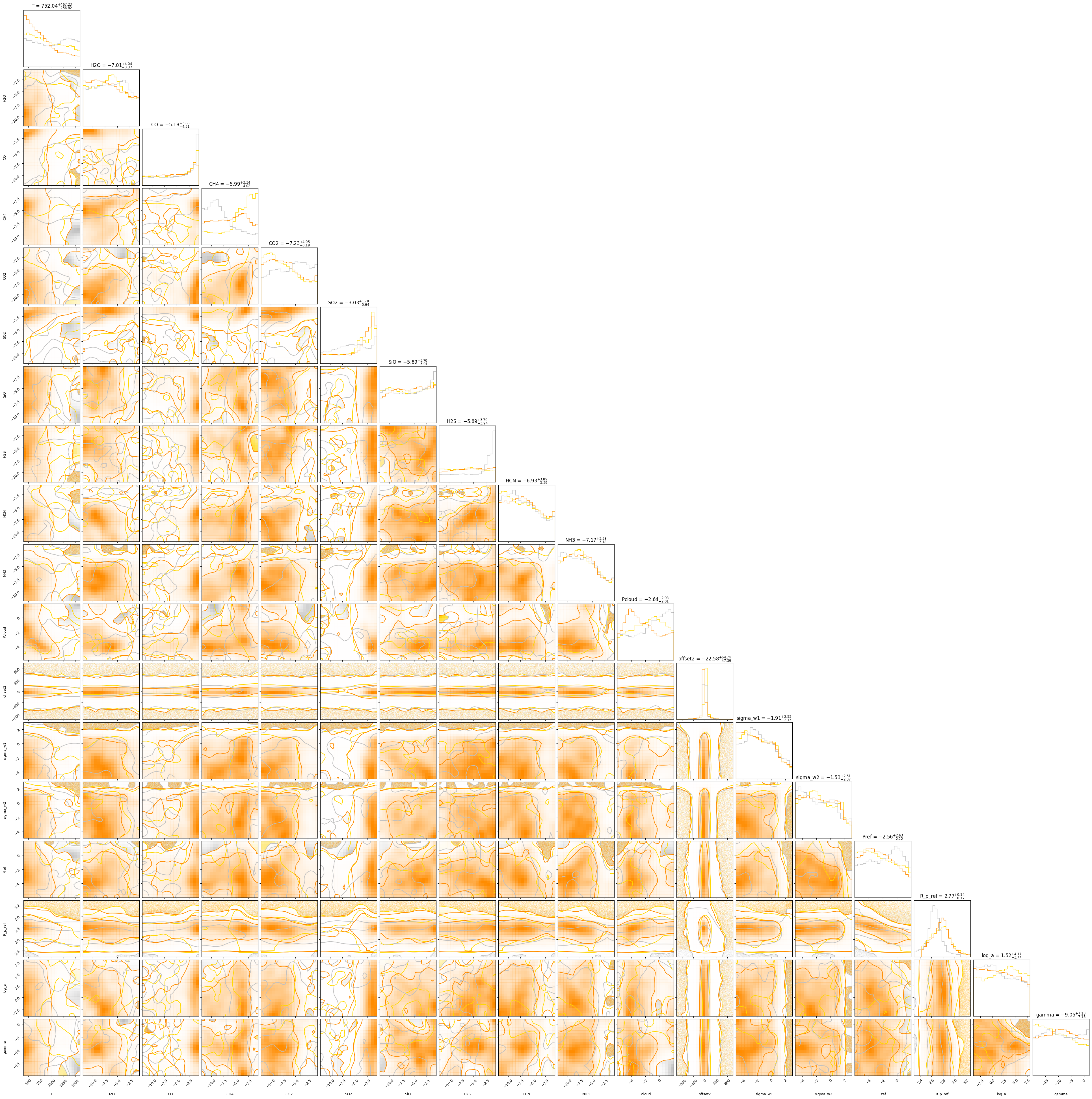}
    \caption{Isothermal, free-chemistry retrieval for TOI-125 c, assuming bulk H$_2$/He atmosphere with a log-uniform prior on the mixing ratio of all trace molecular species. Dark orange indicates posterior results from the \texttt{transitspectroscopy} reduction, yellow from the \texttt{Eureka!} reduction, and grey from the noise spectrum.}
    \label{fig:Cc1}
\end{figure}

\begin{figure}
    \centering
    \includegraphics[width=\linewidth]{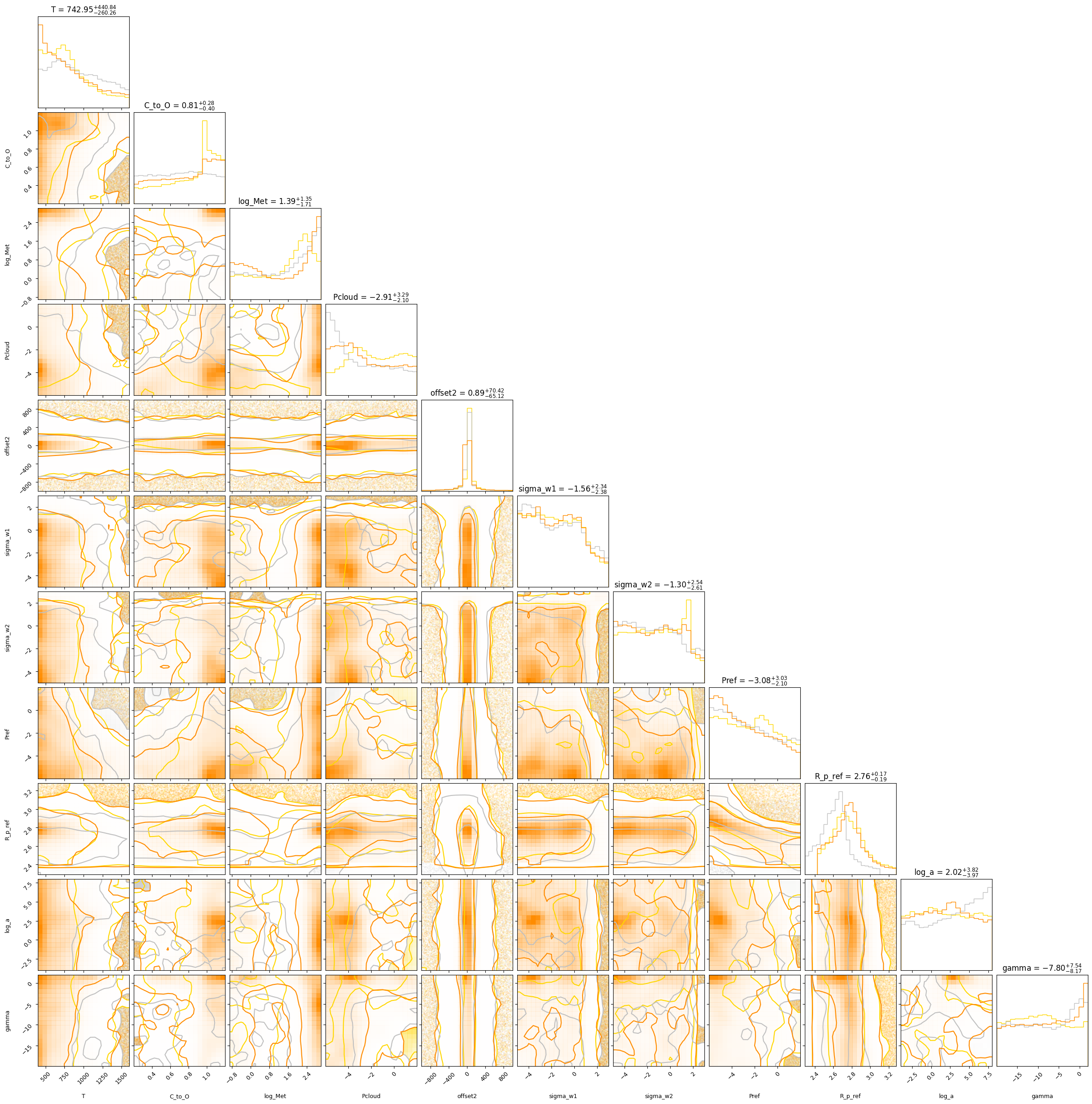}
    \caption{\texttt{POSEIDON} isothermal, chemical equilibrium retrieval for TOI-125 c. Dark orange indicates posterior results from the \texttt{transitspectroscopy} reduction, yellow from the \texttt{Eureka!} reduction, and grey from the noise spectrum.
    }
    \label{fig:Cc3}
\end{figure}

\bibliography{bib}{}
\bibliographystyle{aasjournalv7}

\end{document}